\documentclass[]{interact}

\usepackage{epstopdf}% To incorporate .eps illustrations using PDFLaTeX, etc.
\usepackage[caption=false]{subfig}% Support for small, `sub' figures and tables

\usepackage{bm}
\usepackage{color}
\def\bSig\mathbf{\Sigma}

\usepackage{amsmath}
\usepackage{amssymb}
\usepackage{amsthm}
\usepackage{mathrsfs}
\usepackage{mathtools}

\usepackage{xcolor}
\usepackage{graphicx}
\usepackage{natbib}
\usepackage{booktabs}
\usepackage{amsfonts}
\usepackage{amssymb}
\usepackage{dsfont}
\usepackage{eucal}
\usepackage{multirow}
\usepackage{subfig}

\usepackage{algorithm}
\usepackage{algpseudocode}

\usepackage[font=small,labelfont=bf]{caption}

\usepackage{natbib}% Citation support using natbib.sty
\bibpunct[, ]{(}{)}{;}{a}{}{,}% Citation support using natbib.sty
\renewcommand\bibfont{\fontsize{10}{12}\selectfont}% Bibliography support using natbib.sty

\theoremstyle{plain}% Theorem-like structures provided by amsthm.sty

\theoremstyle{definition}

\theoremstyle{remark}

\DeclareMathOperator*{\argmin}{arg\,min}

\begin{document}

\title{Bayesian Covariate-Dependent Circadian Modeling of Rest-Activity Rhythms}

\author{
\name{Beniamino Hadj-Amar\textsuperscript{a}\thanks{CONTACT: B.H. Email: bh44@rice.edu}, Vaishnav Krishnan\textsuperscript{b} and Marina Vannucci\textsuperscript{a}} 
\affil{\textsuperscript{a}Department of Statistics, Rice University, Houston, TX; \textsuperscript{b}Neurology, Neuroscience, and Psychiatry \& Behavioral Sciences, Baylor College of Medicine, Houston, TX.}
}

\maketitle

\begin{abstract} 
We propose a Bayesian covariate-dependent anti-logistic circadian model for analyzing activity data collected via wrist-worn wearable devices. The proposed approach integrates covariates into the modeling of the amplitude and phase parameters, facilitating cohort-level analysis with enhanced flexibility and interpretability. To promote model sparsity, we employ an $l_1$-ball projection prior, enabling precise control over complexity while identifying significant predictors. We assess performances on simulated data and then apply the method to real-world actigraphy data from people with epilepsy. Our results demonstrate the model's effectiveness in uncovering complex relationships among demographic, psychological, and medical factors influencing rest-activity rhythms, offering insights for personalized clinical assessments and healthcare interventions. \\
	
	\noindent%
	\textbf{Keywords:}  Anti-logistic Circadian Model; Rest-Activity Rhythms;  $l_1$-ball Projection Prior; Multi-subject Modeling;  Wereable Devices.
\end{abstract}

\noindent % 
% {\it Keywords:} Cumulative Shrinkage Prior; Brain Connectivity; Discrete Autoregressive Process; fMRI data; Graphical Models; Horseshoe Prior. 

% \section{Introduction}
% \label{sec:introduction}

% The problem of selecting frequencies that drive the variation in the data is commonly encountered in signal processing, time series analysis, and other fields where understanding the underlying patterns in data is crucial. This problem is often associated with tasks such as feature selection, dimensionality reduction, and model interpretation. Understanding the frequencies that drive the variation in data can improve the interpretability of models. 

% {\color{red} The selection of frequencies is often a complex task and may involve trade-offs between accuracy and computational efficiency.
% Overfitting or underfitting can occur if the wrong frequencies are selected or if too many/few frequencies are considered.}

\section{Introduction}
Recent advancements in wearable technology have revolutionized the monitoring of human activity and rest cycles, offering unprecedented insights into daily fluctuations in a wide variety of physiological parameters \citep{soh2015wearable, chan2012smart}. Among these technologies, actigraphy has received the greatest emphasis, made possible through advances in accelerometers. When worn continuously over several days, wearable accelerometers provide direct measurements of 24h-long behavioral rhythms of rest and activity (rest-activity rhythms, RARs, \citealt{troiano2008physical}). Several FDA-approved research-grade devices are available, capable of measuring one or three-dimensional acceleration at high sampling rates (e.g., 32Hz). Similar accelerometer-based strategies  are now widely integrated into consumer wearables like smart-phones and smart-watches, and can provide reasonably accurate estimates of steps, sleep quality and sedentariness. 

One approach to the analysis of activity data is to model the periodic patterns as a regression problem, with least squares estimators commonly used for inference \citep{cornelissen2014cosinor}. One prominent example is the cosinor model, which employs Fourier series expansions to capture circadian and ultradian rhythms. Bayesian methodologies have extended these models to account for more complex scenarios. For instance, \citet{andrieu1999joint} introduced a Bayesian approach that assumes an unknown number of frequencies, while \citet{hadj2020bayesian, hadj2021identifying} further extended this framework to accommodate nonstationary data. These methods have found extensive applications in fields such as chronotherapy and chronopharmacy, where they are utilized to identify the optimal times for drug administration, striving for maximal efficacy and minimal toxicity \citep{sewlall2010timely, cardinali2021chronotherapy}. % Traditional methods often rely on the cosinor model, which uses Fourier series expansion to model circadian and ultradian rhythms\benni{In the frequentist settings, this is often seen as regression problem where inference is usually carried out via least square estimators \citep{cornelissen2014cosinor}. Bayesian extension include \citet{andrieu1999joint} who assume an unknown number of frequencies and \citet{hadj2020bayesian} who extended to a nonstationary framework. These cosinor methods are widely used in chronotherapy \citep{cardinali2021chronotherapy} and chronopharmacy \citep{sewlall2010timely}, e.g. in many experiments aimed at
% determining the times of highest efficacy and lowest toxicity in response to a variety of
% drugs.} 

Standard cosinor approaches struggle with capturing squared waves, which are common in adult human RARs.  To address this shortcoming, \cite{marler2006sigmoidally} proposed the extended cosinor model, a 5-parameter cosine model that utilizes an anti-logistic function which offers a significant advantage by effectively modeling these complex patterns.  By fitting this model to actigraphy data, researchers have gained insights into individual circadian rhythms, sleep patterns, and overall rest-activity behavior \citep{krafty2019measuring, smagula2022association, abboud2023actigraphic}. Nevertheless, the increased flexibility in modeling the wave signal leads to a more challenging estimation problem due to the presence of nonlinearities and constrained parameters in the model. Moreover, the existing literature on standard and extended cosinor models primarily focuses on modeling medium to large cohorts of individual subjects with parameters calculated separately for each subject. To the best of our knowledge, there is currently no comprehensive modeling approach that models data for an entire cohort of subjects while also incorporating covariates into the framework.

In this article, we propose a Bayesian covariate-dependent anti-logistic circadian (CALC) model for activity data, which simultaneously handles an entire cohort of subjects and incorporates predictors. Specifically, we model the amplitude and phase parameters as functions of the covariates via a log-linear relationship. Previous research has indicated that demographic factors, such as sex and age, are strong determinants of phase and amplitude \citep{eastman2015circadian, mule2021sex, rahman2023age}. In our proposed approach we allow the model to identify the significant effects of predictors on the magnitude of amplitude and phase while promoting sparsity by employing the $l_1$-ball projection prior \citep{xu2023bayesian, hadj2024bayesian}. This prior constrains the regression coefficients within a high-dimensional space bounded by the $l_1$ norm, allowing some coefficients to be exactly zero and thereby identifying the most significant predictor variables. This approach ensures positive probability for coefficients being zero, enabling the calculation of posterior probabilities of inclusion. Furthermore, it is compatible with efficient sampling techniques like Hamiltonian Monte Carlo \citep{duane1987hybrid} and can be easily implemented in probabilistic programming languages such as \textit{stan} \citep{carpenter2017stan}. We propose a methodology for eliciting the $l_1$-ball prior hyperparameters, aiding researchers in achieving desired prior sparsity levels in amplitude and phase modeling. This approach integrates prior information on expected magnitudes, ensuring precise control over model complexity and contributing to advancements in circadian rhythm analysis.

We thoroughly investigate the performance of our proposed methodology through extensive simulation studies, comparing our method with alternative methods. We then illustrate the application of our approach to model actigraphy data from a cohort of adults with focal epilepsy \citep{abboud2023actigraphic}. Our findings underscore the intricate interplay of demographic, psychological, and medical factors in shaping RARs, revealing the complex relationships among parameters influencing rest-activity rhythms and providing valuable insights for personalized healthcare interventions and clinical assessments.

The rest of the paper is structured as follows. Section \ref{sec:methods} introduces the proposed model, including prior formulation, posterior inference and our approach for eliciting the prior. Section \ref{sec:simul_study} presents the results from simulation studies, while Section \ref{sec:application} demonstrates the application of our methodology to actigraph data. Finally, Section \ref{sec:concluding_remarks} offers some concluding remarks. Stan files and R utilities are available in Github at XXX (to be made public upon acceptance).
%\url{https://github.com/Beniamino92/CALC}

% \benni{A rich literature exists for both discrete and continuous shrinkage estimators/priors in linear models, the most pop- ular being the Lasso method (Tibshirani 1996; Park and Casella 2008). Such models shrink unrestricted coefficients in an attempt to select among many competing predictors.The Bayesian variable selection literature also includes stochas- tic search or “spike-and-slab” priors, typically characterized by two-component mixture priors for coefficients allowing for “on” and “off” settings (George and McCulloch 1993, 1997). somewhere cite \citet{hadj2023bayesian} }

% \benni{here I need to say that $l_1$ ball does not need to do stocasthic search which in high dimension can be a mess, but it just projects. }

% \benni{What does our model help us with?. This integration of sophisticated statistical techniques ensures more accurate and personalized treatment plans, enhancing patient outcomes.}

\section{Methods} \label{sec:methods}
% Let $y_{it}$ denote the log activity count of individual $i$ at time $t$, for $t = 1, \dots, T_i$. As commonly done with log count data, a one is added to each count before performing the logarithmic operation, to prevent taking the logarithm of zero counts. We assume data collected from $N$ individuals and allow individuals to have varying recording times, $T_i$, based on their availability to participate in the experiment.
% %where each subject, indexed as $i \in \{1, \dots, N\}$,  has recorded activity levels over $T_i$ time points, with $T_i$ being the length of time series for subject $i$. This reflects that individuals may have varying recording times based on their availability to participate in the experiment.
% Additionally, let $\bm{x}_i = (x_{i1}, \dots, x_{iQ})'$ represent the values of $Q$ covariates for individual $i$, where covariates are assumed to be both discrete and continuous; without loss of generality, we assume an intercept $x_{i1} = 1$. 

 Let \(Y_{it}\) be the outcome of interest at time \(t\) for individual \(i\), for $i=1, \ldots, N$ and $t =1, \ldots, T_i$,  with $T_i$ indicating the number of observations for the $i$-th individual. 
	%We consider a total of \(N\) individuals, each with \(T_i\) observations reflecting differences in data collection windows. 
	%Our proposed framework accommodates various outcome types, including continuous data or transformations of count data. 
	In the present application, we define \(y_{it} = \log(1 + Y_{it})\), with $Y_{it}$ an observed count, as this log-count transformation is commonly used in actigraphy studies to mitigate zero inflation and stabilize variance. However, our proposed methodology is not restricted to this particular transformation and can be applied to other outcome types. For each individual \(i\), let \(\bm{x}_i = (x_{i1}, \dots, x_{iQ})'\) denote the values of \(Q\) covariates, which may be discrete or continuous. If needed, and without loss of generality, one may include an intercept by setting \(x_{i1} = 1\).

\subsection{Covariate-dependent multi-subject anti-logistic circadian model}
\label{sec:model}
As in \citet{krafty2019measuring}, we assume that activity level is recorded within adjacent and disjoint epochs of length 1/$R$ hours; or, in an equivalent way, we assume that the activity counts are recorded in such a way that the sampling rate corresponds to $R$ intervals per hour. In our application, activity is considered within 5 minutes intervals, so that $R = 12$. 
We model the log-activity counts $y_{it}$ through a covariate-dependent anti-logistic circadian model (CALC) of the form 
\begin{equation}
	y_{it} =  m_i + a(\bm{x}_i, \bm{\eta}_a) \text{expit} \Big\{ \beta_i \big[ \cos{\bigg(\frac{t}{R} - \phi(\bm{x}_i, \bm{\eta}_\phi) \bigg) \frac{2\pi}{24}} \big] - \alpha_i \Big\} + \varepsilon_{it},
	\label{eq:multisubj_RAR_model}
\end{equation}
where $\varepsilon_{it} \sim \mathcal{N}(0, \sigma^2_i)$, for $i = 1, \dots, N$, and $t = 1, \dots, T_i$, and where $\text{expit}(\theta) = \frac{\exp{ \{\theta\}}}{(1 + \exp{ \{\theta\}})}$ is the \textit{anti-logistic}, or expit, function \citep{marler2006sigmoidally}. The parameter $m_i$ represents the minimum expected  activity level, while $\beta_i$ serves as the shape parameter, which is directly related to the rate of change during the transition from rest to activity, with larger values indicating faster transitions. The parameter $\alpha_i$ governs the duration of rest compared to active time, with higher positive values suggesting less active time. The covariate-dependent parameters $a(\bm{x}_i, \bm{\eta}_a)$ and $\phi(\bm{x}_i, \bm{\eta}_\phi)$ represent the amplitude, namely the difference between maximum and minimum values, and the acrophase, which denotes the time of day when peak activity occurs, respectively.  For identifiability, we assume the following constraints for the modeling parameters: $m_i>0$, $a(\bm{x}_i, \bm{\eta}_a)>0$, $\alpha_i \in (-1, 1)$, $\beta_i > 0$, and $\phi(\bm{x}_i, \bm{\eta}_\phi) \in [0, 24]$, for $i = 1, \dots, N$.  

\begin{figure}[htbp]
	\centering
	\includegraphics[width=0.8\textwidth]{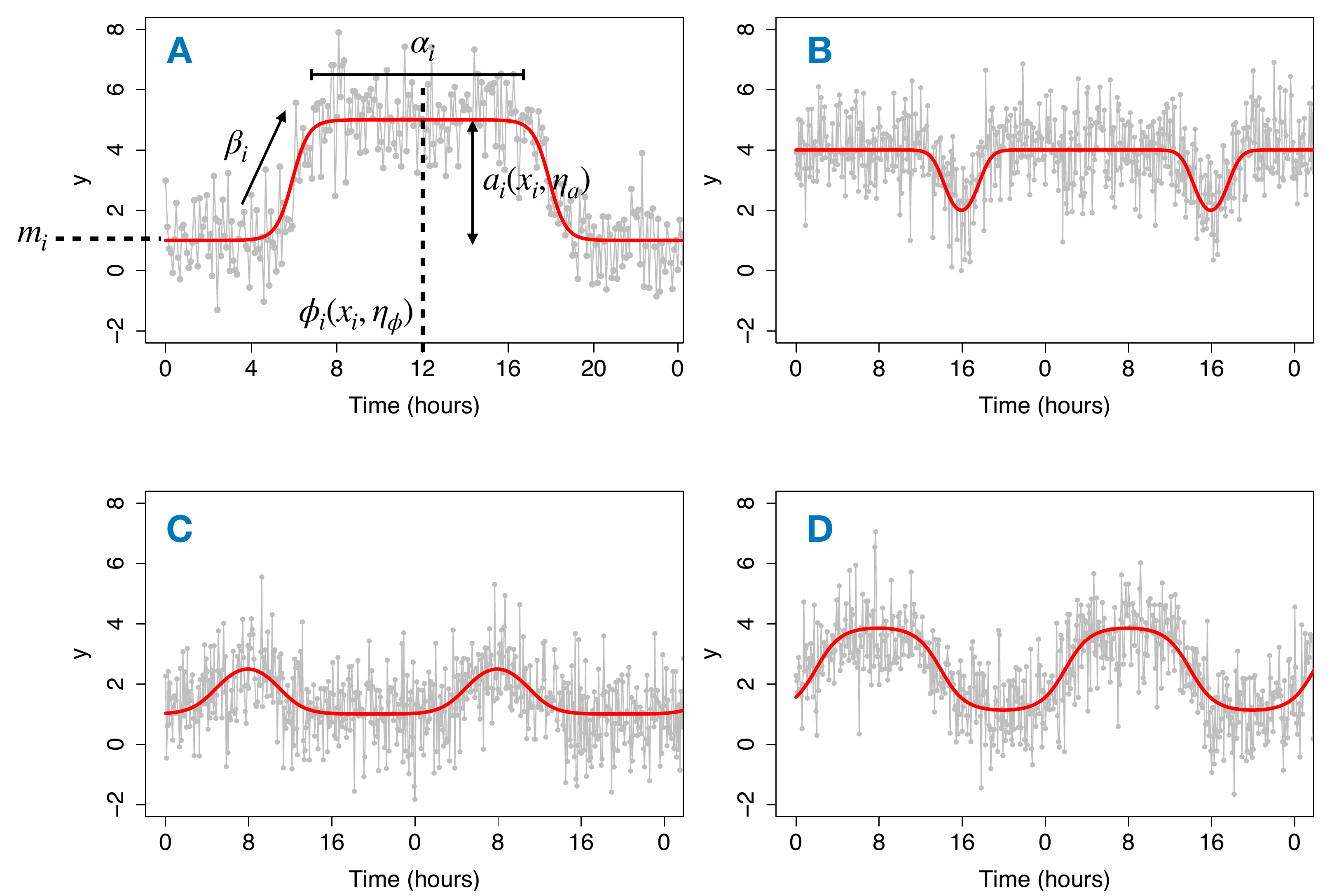}
	\caption{Simulated realizations (dots) and true generating signal (red line) for four different combinations of the modeling parameters: (a) $m_i = 1, a(\cdot) = 4, \alpha_i=0, \beta_i = 10$, and $\phi(\cdot) = 12$; (b) $m_i = 0, a(\cdot) = 4, \alpha_i=-0.99, \beta_i = 10$, and $\phi(\cdot) = 4$; (c) $m_i = 1, a(\cdot) = 3, \alpha_i=0.99, \beta_i = 3$, and $\phi(\cdot) = 8$; (d) $m_i = 1, a(\cdot) = 3, \alpha_i=0, \beta_i = 3$, and $\phi(\cdot) = 8$; for all scenarios $\sigma^2_i = 1$.}
	\label{fig:intro_plot}
\end{figure}

We provide an intuitive illustration of the proposed model \eqref{eq:multisubj_RAR_model} in Figure \ref{fig:intro_plot}, which 
shows simulated realizations and generating signal for four different combinations of the modeling parameters: (a) $m_i = 1, a(\cdot) = 4, \alpha_i=0, \beta_i = 10$, and $\phi(\cdot) = 12$; (b) $m_i = 0, a(\cdot) = 4, \alpha_i=-0.99, \beta_i = 10$, and $\phi(\cdot) = 4$; (c) $m_i = 1, a(\cdot) = 3, \alpha_i=0.99, \beta_i = 3$, and $\phi(\cdot) = 8$; (d) $m_i = 1, a(\cdot) = 3, \alpha_i=0, \beta_i = 3$, and $\phi(\cdot) = 8$; for all scenarios $\sigma^2_i=1$. From these illustrations we can observe how different combinations of model parameters give rise to various curve shapes, potentially accommodating diverse patterns of rest-activity rhythms.

Next, we model the amplitude and phase parameters as functions of covariates by assuming that their logarithms are linear combinations of the covariates, $\bm{x}_i$, with linear coefficients $\bm{\eta}_a$ and $\bm{\eta}_{\phi}$, respectively, as

\begin{equation}
	a(\bm{x}_i, \bm{\eta}_a)  = \exp(\bm{x}'_i \bm{\eta}_a), \qquad 
	\phi(\bm{x}_i, \bm{\eta}_\phi) = \exp(\bm{x}'_i \bm{\eta}_{\phi}), 
\end{equation}
where $\bm{\eta}_{a} = (\eta_{a,1} \ldots, \eta_{a,Q})$ and $\bm{\eta}_{\phi} = (\eta_{\phi,1} \ldots, \eta_{\phi,Q})$. This enables the discovery of potential relationships within the data. Our decision to model only the amplitude and phase as functions of covariates was informed by previous research indicating that covariates such as sex, age and other demographic factors play a significant role in influencing these parameters \citep{eastman2015circadian, mule2021sex, rahman2023age}. We note that, while it is true that conventional cosinor models have revealed significant distinctions in the midline estimating statistics of rhythm (MESOR), which may be erroneously identified with $m_i$, in our proposed model $m_i$ assumes a distinct interpretation, representing the minimum value of the curve rather than the midline value. Finally, modeling all parameters in the model as functions of the covariates would increase the computational complexity considerably, making inference within this framework generally not straightforward due to evident complex nonlinearities, such as the presence of cosine within an anti-logistic function.

% \begin{figure}[htbp]
%     \centering
%     \includegraphics[width=0.8\textwidth]{figures/plot_intro_final.pdf}
%     \caption{\benni{something}}
%     \label{fig:intro_plot}
% \end{figure}

\subsection{Promoting sparsity via the $l_1$-ball projection prior}
\label{sec:l1ball}

A sparse solution is essential for identifying the subset of predictor variables $\bm{x}$ that have a significant impact on the level of activity $y_{it}$. Hence, we enforce sparsity on the linear parameters $\bm{\eta}_{a}$ and $\bm{\eta}_{\phi}$, determining the magnitude of amplitude and phase, respectively, by employing the $l_1$-ball projection prior \citep{xu2023bayesian}. 
%Recently,  \citet{hadj2024bayesian} employed this prior for promoting sparsity of the vector-autoregressive emission coefficients  within a hidden semi-Markov model framework.  
The $l_1$-ball refers to a geometric shape in high-dimensional space consisting of all points whose $l_1$ norm is bounded by a radius $r$. In this context, the $l_1$ ball projection prior constrains the coefficients of the regression model to lie within this geometric shape, while being able to assign mass specifically at exact zeros. By using such a prior, we introduce unconstrained latent variables and then transform them onto the space of constrained parameters. This transformation ensures that there is positive probability assigned to any element being exactly zero, hence enabling us to estimate posterior probabilities of inclusion. 
Alternative options for variable selection, such as spike and slab priors \citep{mitchell1988bayesian, george1997approaches, Brown1998, ishwaran2005spike}, that employ a point mass distribution at zero, entail the exploration of the space of possible models by selecting subsets of the variables, each associated to the value of a vector of binary indicators. See \cite{Vannucci2021} for a comprehensive treatment of these priors. On the contrary, the $l_1$-ball projection algorithm changes a combinatorial problem of choosing which subset of the parameter is zero into a continuous optimization problem, yielding computational and modeling gains.

Formally, define $\bm{\psi}_{z} = (\psi_{z,1} \dots, \psi_{z,Q})$, for $z \in \{a, \phi \}$, the vector of latent parameters associated to amplitude and phase, respectively; similarly let $\bm{\eta}_{z} = (\eta_{z,1} \dots, \eta_{z,Q})$ the vectors of linear coefficients.  A double-exponential distribution is assumed on random variables $\psi_{z,l}$, and the $l_1$-ball projection with radius $r_z$ is applied to map $\bm{\psi}_{z} \rightarrow \bm{\eta}_z$ in such a way that if $|| \bm{\psi}_{z} || \geq r_z $, then $||\bm{\eta}_z || = r_z$ with some of the $\eta_{z,l} = 0$. That is,  
\begin{equation}
	\begin{split}
		\big\{ \psi_{z,l} \big\}_{l = 1}^{Q} \overset{\mathrm{iid}}{\sim} &\text{DExp} (0, \tau_z), \quad  r_z \sim \text{Expo}(\lambda_{x}) \\ 
		&\bm{\eta}_{\,z} := \argmin_{||\bm{s}||_1 \, \leq \, r_{\,z}} ||\bm{\psi}_{x} - \bm{s}||_2^2, 
	\end{split}
	\label{eq:l1ball}
\end{equation}
where $||\cdot||_1$ and $||\cdot||_2$ denote $l_1$ and $l_2$ norms, respectively, and where the radius $r_z > 0$. Here, Expo$(\lambda_z)$ represents an exponential distribution with rate $\lambda_z$, and $\text{DExp}(0, \tau_z)$ denotes a double-exponential distribution with mean 0 and scale parameter $\tau_z$. The loss function in Eq. \eqref{eq:l1ball}
is strictly convex, namely for every $\bm{\psi}_{\,z}$, there is only one optimal solution $\bm{\eta}_{\,z}$. Crucially, this transformation is almost surely continuous and differentiable,  with its Jacobian being equal to one.  Furthermore, this approach remains compatible with efficient posterior sampling techniques such as Hamiltonian Monte Carlo (HMC; \citealt{duane1987hybrid}) and widely-used probabilistic programming languages like \textit{stan} \citep{carpenter2017stan}. Note that even though the prior on the latent variable \( \psi_{z,l} \) only needs to be continuous, we opt for a double exponential distribution. This choice enables us to differentiate the prior specification for \( \bm{\eta}_z \) from the prior specification on \( r_z \), as detailed in Section \ref{sec:hyperparms_elictation}, thus facilitating the selection of hyperparameters in a principled manner.
To complete our prior specification for the model parameters in Eq. \eqref{eq:multisubj_RAR_model}, we apply a log-scale transformation to the expected activity levels $m_i$ and shape parameters $\beta_i$, ensuring positive support and interpretability. Specifically, we set $m_i = \exp{\{lm_i\}}$ and $\beta_i = \exp{\{l\beta_i\}}$, where $lm_i \sim \mathcal{N}(l\mu_m, \sigma^2_m)$ and $l\beta_i \sim \mathcal{N}(l\mu_\beta, \sigma^2_\beta)$. This formulation centers the parameters around population means ($\mu_m$ and $\mu_\beta$) while allowing individual-level variability. We further adopt a partial pooling strategy for the hyperparameters, specifying diffuse prior distributions to account for uncertainty in the population-level parameters. In particular, we assume $l\mu_m \sim \mathcal{N}(0, 10)$ and $l\mu_\beta \sim \mathcal{N}(5, 3)$, enabling data-driven estimation of the population means while leveraging information across subjects to improve individual parameter estimates. Finally, we place priors on the remaining model parameters. The parameter $\alpha_i$, governing the duration of rest compared to
	active time, is assigned a Beta distribution, $\alpha_i \sim \text{Beta}(a, b)$, with $\text{Beta}(1,1)$ recovering the $\text{Unif}(0,1)$ prior as a special case. The residual scale parameter $\sigma_i$ follows a half-Cauchy distribution, $\sigma_i \sim C^+(0, \gamma_\sigma)$, where $C^+$ denotes the truncated Cauchy distribution with support on the positive real line. This choice of priors ensures both flexibility and regularization in the model's parameter estimation.

\subsection{Posterior Inference}
\label{sec:posterior_inference}
Let $\bm{\theta} = (\bm{m}', \bm{\alpha}', \bm{\beta}', \bm{\sigma}', \bm{\psi}'_a, \bm{\psi}'_\phi)'$ denote the set of parameters to be inferred, with $\bm{m} = (m_1, \dots, m_N)'$ the vector of minima, $\bm{\alpha} = (\alpha_1, \dots, \alpha_N)'$ the vector of parameters governing the duration of rest compared to active time, $\bm{\beta} = (\beta_1, \dots, \beta_N)'$ the vector of shape parameters, $\bm{\sigma} = (\sigma_1, \dots, \sigma_N)'$ the vector of standard deviations, and $\bm{\psi}_a$ and $\bm{\psi}_\phi$ the vectors of latent (unconstrained) parameters associated with amplitude and phase, as defined in Section \ref{sec:l1ball}. We remark that the linear parameters $\bm{\eta}_a$ and $\bm{\eta}_\phi$ are constructed deterministically from  $\bm{\psi}_a$ and $\bm{\psi}_\phi$. Furthermore, let us define $\bm{X} = \{ \bm{x}_i \}_{i=1}^N$ as the collection of covariates corresponding to all subjects. Then, the posterior distribution can be expressed as 
\begin{equation}
	p (\bm{\theta} | \bm{y}, \bm{X}) \propto \mathscr{L}\,  ( \bm{y} | \bm{\theta}, \bm{X}) \times p (\bm{\psi}_a) \times  p (\bm{\psi}_\phi)
	\times \prod_{i=1}^N  \bigg\{ p (m_i)  \times p (\alpha_i) \times  p (\beta_i) \times p (\sigma_i)  \bigg\},
	\label{eq:posterior}
\end{equation}
where the likelihood $\mathscr{L}\,  ( \cdot )$ is defined as 
\begin{equation*}
	\mathscr{L}\,  ( \bm{y} | \bm{\theta}, \bm{X}) = \prod_{i=1}^{N}\prod_{t = 1}^{T_i} \mathcal{N} \Bigg(m_i + a_i(\bm{x}_i, \bm{\eta}_a) \text{expit} \Big\{ \beta_i \big[ \cos{\bigg(\frac{t}{R} - \phi_i(\bm{x}_i, \bm{\eta}_\phi) \bigg) \frac{2\pi}{24}} \big] - \alpha_i \Big\}, \sigma^2_i \Bigg).
	\label{eq:likelihood}
\end{equation*}

The posterior distribution \eqref{eq:posterior} is not analytically tractable, leading us to employ MCMC methods for inference. Our approach relies on Hamiltonian Monte Carlo \citep{duane1987hybrid}, which leverages a discretized version of Hamiltonian dynamics to propose joint parameter updates. These updates are evaluated using the standard Metropolis-Hastings probability framework, and the No-U-Turn Sampler (NUTS) \citep{hoffman2014no}, which dynamically adjusts the Hamiltonian step size to match the posterior's geometry. This combined strategy facilitates efficient exploration of the multivariate parameter space. Leveraging the capabilities of the probabilistic programming language \textit{stan} \citep{carpenter2017stan}, we can seamlessly implement HMC and NUTS. For this model, \textit{stan} simplifies the process by only requiring users to specify the priors, the algorithm used to solve the projection of Eq. \eqref{eq:l1ball}, and the necessary computations for likelihood evaluation. It then handles model compilation and optimizes sampling according to the model's geometry.  Note that constraints are met in \textit{stan} via both parameters and transformed parameters blocks, ensuring that proposed values are rejected if the transformed parameter fails to align with its constraint (e.g., phase $\phi(\bm{x}_i, \bm{\eta}_\phi)$ within the range [0, 24]). Additionally, compilation and optimization of the sampler is aided by ensuring that priors are appropriately specified as not excessively diffused. The projection of Eq. \eqref{eq:l1ball} can be solved using the procedure outlined in \citet{xu2023bayesian}, summarized in Algorithm \eqref{Alg:l1_ball} below. 

\begin{algorithm}[htbp]
	\caption{$l_1$-ball projection. $\, \, $ \textit{\textbf{Input:}}
		$\bm{\psi}_z \in \mathbb{R}$ and $r_{\,z} \in \mathbb{R}_{+}$. $\, \, $ \textit{\textbf{Output:}}
		$\bm{\eta}_z \in \mathbb{R}^{D} \cup \{0\}$ }\label{alg:l1_balla}
	\begin{algorithmic}
		\If{$||\bm{\psi}_z||_1 \leq r_{z}$}
		\State $\bm{\eta}_z \gets \bm{\psi}_z$
		\Else    
		\State Sort $\bm{\psi}_z$ so that $|\psi_{z\,(1)}| \geq \cdots \geq |\psi_{z\,(Q)}|$ 
		\State $\phi_l\gets \left(\sum_{n=1}^l |\psi_{z\,(n)}| - r_{z}\right)_{+},$ $\, \qquad \quad \text{for } \, l = 1, \dots, Q. $ 
		\State $m \gets \max\left\{n: |\psi_{z\,(n)}| > \frac{\phi_n}{n}\right\}$
		\State $\tilde{\phi} \gets \frac{\phi_m}{m}$
		\State $\eta_{z,i} \gets~ \textup{sign}(\psi_{z,i})\, \max\left(|\psi_{z,i}| - \tilde{\phi}, 0 \right),$ $\, \, \text{for } \, l = 1, \dots, Q. $ 
		%\State $\bm{\Theta}_p^{\,j} \gets vec^{-1}(\bm{\theta}_p^{\,j})$
		\EndIf
	\end{algorithmic}
	\label{Alg:l1_ball}
\end{algorithm}

%\times \prod_{i=1}^N   p (m_i) \times p (\alpha_i) p (\alpha_i) p (\beta_i) \times p (\sigma^_i)

\subsection{$l_1$-ball hyperparameter elicitation}
\label{sec:hyperparms_elictation}
We outline the methodology for determining suitable values for prior hyperparameters $\tau_{z}$ and $\lambda_{z}$, necessary to complete the $l_1$-ball prior elicitation for $z \in \{a, \phi \}$. \citet{xu2023bayesian} demonstrated that the prior specification delineated in Equation \eqref{eq:l1ball} results in a double exponential distribution for the induced prior on active $\eta_{z,l}$, expressed as $\pi(\eta_{z,l} | \eta_{z, l} \neq 0) \sim \text{DExp}(0, \tau_z)$. Consequently, as highlighted in \citet{hadj2024bayesian}, this property facilitates the direct sampling from the induced prior distribution on $\eta_{z,l}$ to achieve a preferred prior expected sparsity level, say $\xi_z \in (0, 1)$, for $z \in \{a, \phi\}$. All that is required is to simulate, using a Monte Carlo approach, from a grid of hyperparameter values for the $l_1$ ball prior and select the one closest to the expected sparsity. For instance, in our simulation study, we set $\tau_z = \lambda_z = 1$, ensuring an expected sparsity of approximately 90\%, for both amplitude and phase. 

Alternatively, if a researcher has prior information about amplitude and phase, we propose to integrate this prior information in a more informative manner into the model. This is achieved by integrating prior knowledge concerning the upper bounds of expected magnitudes for both amplitude and phase, denoted as $q_a$ and $q_\phi$ respectively. For instance, a medical researcher might posit that the average amplitude of individuals does not exceed $q_a = 8$, expressing a confidence level in this assertion, say $p_a = 0.95$. Analogously, a similar assertion could be made regarding phase. Consequently, we propose to empower decision-makers by formulating the hyperprior specification in a manner that ensures the prior probability of the average magnitude of $a(\cdot, \cdot)$ and $\phi(\cdot, \cdot)$  not exceeding $q_z$ is governed by a predetermined probability $p_z$. This can be expressed as: 
\begin{equation}
	\dfrac{1}{N} \sum_{i=1}^N p \bigg( a(\bm{x}_i, \bm{\eta}_a) \leq q_a \bigg) = p_a, \qquad   \dfrac{1}{N} \sum_{i=1}^N p \bigg( \phi(\bm{x}_i, \bm{\eta}_\phi) \leq q_\phi \bigg) = p_\phi.
	\label{eq:magnitude}
\end{equation}
Given predetermined values for $\xi_z$, $q_z$, and $p_z$, where $z \in \{a, \phi\}$, we propose a two step procedure to elicit prior hyperparameters $\sigma_{z}$ and $\lambda_{z}$ through Monte Carlo (MC) simulation over $S$ iterations. We first elicit $\tau_{z}$ by constructing a one-dimensional grid comprising potential candidate values $\bm{\tau}_z = (\tau_{z, 1}, \dots, \tau_{z, J})$. For each candidate $\tau_{z,j}$, at MC iteration $s$, we  draw $\eta^{s}_{z,l} \sim \xi_z \delta_0 + (1-\xi_z) \text{DExp}(0, \tau_{z, j})$, for $l=1, \dots, Q$,  and compute the corresponding $p_z$th percentile $q_{z, j}^{s} $ for different subjects using Eq. \eqref{eq:magnitude}. These values, $q_{z, j}^{s} $, are then averaged across MC iterations to derive $\hat{q}_{z,j}$. Finally, $\hat{\tau}_{z}$ is selected by minimizing the absolute error between the amplitude and the expected upper bound, expressed as $ \hat{\tau}_z = \argmin_{\tau_{z,j}} \big\{  \big| \hat{q}_{z, j} - q_z \big| \big\}$.  
Once $\hat{\tau}_z$ has been determined, we revert to considering $\bm{\eta}_z$ as a function of $\bm{\psi}_z$ and $r_z$, and set $\lambda_z$ such that the prior expected sparsity of the vector $\bm{\eta}_z$ is $\xi_z$, expressed as:
\begin{equation*}
	\mathbb{E}_{r_z \sim \text{Exp}(\lambda_z}) \mathbb{E}_{\bm{\psi}_x \sim \text{DExp}(0, \hat{\tau}_z)} \bigg[\dfrac{1}{Q} \sum_{l=1}^{Q} \mathds{1}_{\eta_{z,l} = 0}  \bigg] = \xi_z
\end{equation*}
This is achieved once more through Monte Carlo estimation, exploring a grid of values $\bm{\lambda}_z = \{\lambda_{z, 1}, \dots, \lambda_{z, J} \}$. Initially, we draw $\bm{\eta}^{s}_z \sim l_1\text{-ball}(\hat{\tau}_z, \lambda_{z,j})$, employing the generative model described in Equation \eqref{eq:l1ball} and the projection algorithm outlined in Equation \eqref{Alg:l1_ball}. Subsequently, we estimate the induced sparsity level as $\hat{\xi}_{z,j} = \frac{1}{S}\frac{1}{Q}\sum_{s=1}^{S}\sum_{l=1}^{Q}  \mathds{1}_{{\eta^{s}_{z,l}}} (0)$. The pseudo-code of this procedure is given in Algorithm \eqref{alg:hyperparms_elictation}, below. 
\begin{algorithm}[htbp]
	\small
	\caption{Hyperameters Elicitation, $\quad z \in \{a, \phi\}$}\label{alg:hyperparms_elictation}
	\begin{algorithmic}

		\State \textit{Scale Double Exponential}: DExp$(0, \tau_z)$
		\For{$\tau_{z,j} \in  \bm{\tau}_z$}        
		\For{$ s = 1, \dots, S$}
		\State Draw $\eta^{s}_{z,l} \sim \xi_z \delta_0 + (1-\xi_z) \text{DExp}(0, \tau_{z, j})$, \quad $l = 1, \dots, Q.$
		% \State Compute $ a(\bm{x}_i, \bm{\eta}_x^{s})_{\sigma_{x,j}}, \quad \for i = 1, \dots, N. $
		\State Compute $q_{z,j}^{s}$ s.t. $\frac{1}{N} \sum_{i=1}^N p \bigg( a(\bm{z}_i, \bm{\eta}^{s}_z) \leq q_{z,j}^{s} \bigg) = p_z.$
		\EndFor 
		\EndFor
		\State Set $\hat{\tau}_z = \argmin_{\tau_{z,j}} \big\{  \big| \hat{q}_{z, j} - q_z \big| \big\}$, \quad where $\hat{q}_{z,j} = \frac{1}{S}\sum_{s=1}^S q_{z, j}^{s}$.
		\State
		
		\State {\textit{Rate Exponential}: Expo$(\lambda_z)$}
		\For{$\lambda_{z,j} \in  \bm{\lambda}_z$}        
		\For{$ s = 1, \dots, S$}
		\State Draw $\bm{\eta}^{s}_{z} \sim l_1$-ball $(\hat{\tau}_z, \lambda_{z,j})$ via Algorithm \eqref{Alg:l1_ball}
		% \State Compute $ a(\bm{x}_i, \bm{\eta}_x^{s})_{\sigma_{x,j}}, \quad \for i = 1, \dots, N. $
		\EndFor 
		\State Compute $\hat{\xi}_{z,j} = \frac{1}{S}\frac{1}{Q}\sum_{s=1}^{S}\sum_{l=1}^{Q} \{ \mathds{1}_{\eta^{s}_{z,l} = 0} \}$
		\EndFor
		\State Set $ \hat{\lambda}_z = \argmin_{\hat{\lambda}_{z,j}} \big\{  \big| \xi_{z,j} - \xi_z \big| \big\} $
	\end{algorithmic}
	\label{Alg:hyperparms_elictation}
\end{algorithm}

% \caption{Hyperameters Elictation, $\quad x \in \{a, \phi\}$ $\, \, $ \\ \textit{\textbf{Input:}}
% $\xi_x \in (0, 1)$, $q_{\,x} \in \mathbb{R}_{+}$, p_{\,x} \in (0, 1)$\, \, $  \\ \textit{\textbf{Output:}}
% $ \sigma_x, \lambda_x $ }

\section{Simulation Study}
\label{sec:simul_study}

We assess performances of our proposed method through simulated data and compare results with alternative methods.

\subsection{Data generation and parameter setting}
\begin{figure}[htbp]
	\centering
	\includegraphics[width=0.8\textwidth]{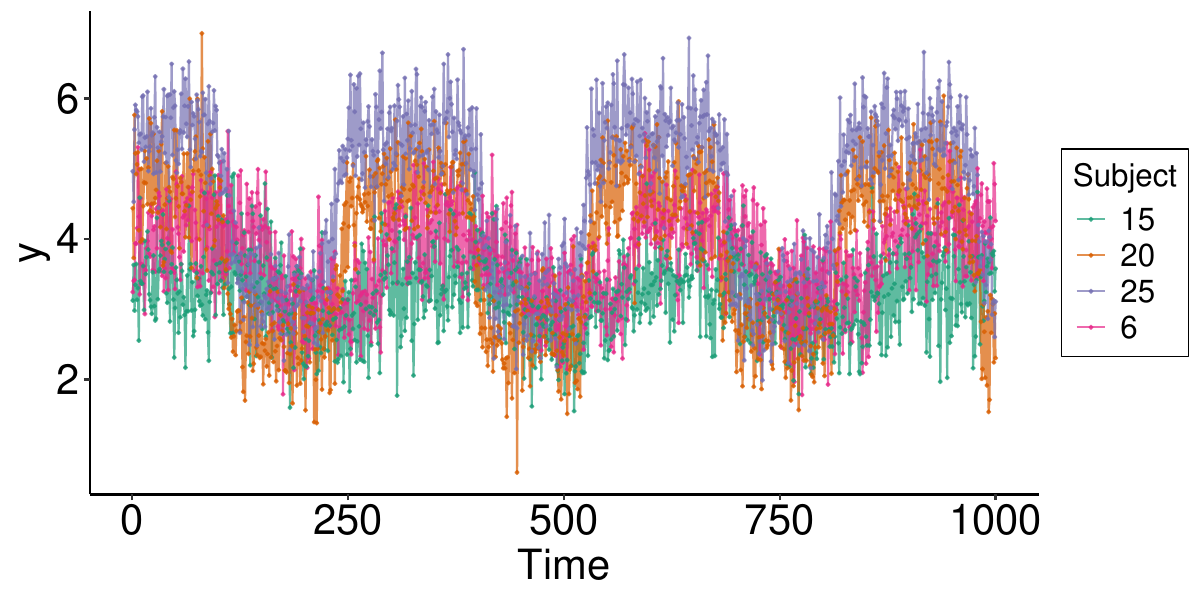}
	\caption{Simulation Study.  Simulated time series data for four subjects.  }
	\label{fig:simul_data}
\end{figure}

We generated time series data from a cohort of $N=30$ individuals, with each individual contributing $T = 1000$ observations, resulting in a collective dataset of 30,000 data points. Our analysis consider $R = 12$. Covariates were incorporated to influence both amplitude and phase parameters. With $Q = 15$ covariates considered for each subject, our design encompassed a mix of continuous and discrete features. Covariates were generated as follows: for the first fifteen subjects, $x_{i1}$ was set to 1, while for the subsequent fifteen subjects, $x_{i1}$ was set to 0, noting that an intercept term was not included in the simulation study. For $x_{i2}$ and $x_{i3}$, half of the values were sampled from a mixture of normal distributions with means 180 and 160 (for $x_{i2}$) and 75 and 55 (for $x_{i3}$), each with a variance of 2. Covariates $x_{i7}$ and $x_{i,10}$ were generated from Bernoulli distributions with parameter 0.5, while the remaining covariates were sampled from standard normal distributions. Subsequently, the continuous covariates were standardized such that each column of the design matrix had a mean of zero and a variance of one.   Linear coefficients governing amplitude $\bm{\eta}_{a}$ were  assigned values of $\eta_{a,1} = 0.6$, $\eta_{a,4} = 0.5$, $\eta_{a,5} = -0.6$, $\eta_{a,7} = 0.1$ and $\eta_{a,l} = 0$ for $l \notin \{1, 4, 5, 7\}$. These values cover both positive and negative amplitude effects, with $\eta_{a,7} = 0.1$ representing a notably small effect. Linear coefficients regulating the covariate-dependent phase  $\bm{\eta}_{\phi}$ were set to $\eta_{\phi,1} = 1.0$, $\eta_{\phi,2} = 0.9$, and a small effect $\eta_{\phi,3} = 0.01$, while all other coefficients $\eta_{\phi,l}$ were set to zero, for $l \notin \{1, 2, 3\}$. We observe that, for this example, the solution for the covariate-dependent phase and amplitude exhibits considerable sparsity. The remaining subject-specific  parameters for data generation  were simulated as $m_i \sim \mathcal{N}(3, 0.5)$, $\beta_i \sim \mathcal{N}(10, 0.5)$, and $\alpha_i \sim \mathcal{N}(-0.3, 0.1)$, allowing for individual heterogeneity, while we set $\sigma_i = 0.5$, for $i = 1, \dots, N$.  Figure \ref{fig:simul_data} displays simulated time series data for four subjects, illustrating two distinct patterns in both amplitude and phase behavior across the corresponding time series data.

In this scenario, we aim to simulate a situation where prior information on the magnitude of amplitude and phase is absent, yet we anticipate a certain level of sparsity. Accordingly, we chose the hyperparameters for the $l_1$-ball prior to embody a prior sparsity assumption of 90\%, thus setting $\lambda_a = \lambda_\phi = 1$ for the exponential prior on the ray and $\tau_a = \tau_\phi = 1$ for the double-exponential prior on the latent coefficients.  In finalizing our hyperparameter specification for the model parameters, we specified $\sigma_i \sim C^+(0, 1)$, weakly informative, for $i = 1, \dots, N$. 

MCMC chains were run for 3,000 iterations, excluding the first 1,000 as the burn-in period, leaving 2,000 for inference. The convergence of the MCMC chains was evaluated using the Gelman-Rubin convergence diagnostic R-hat, yielding statistics with values of $R$ not exceeding 1.01, hence showing no pathological behavior. We provide representative trace plots of the parameters in the Appendix. 

\subsection{Results}

\begin{figure}[htbp]
	\centering
	\includegraphics[width=.6\linewidth]{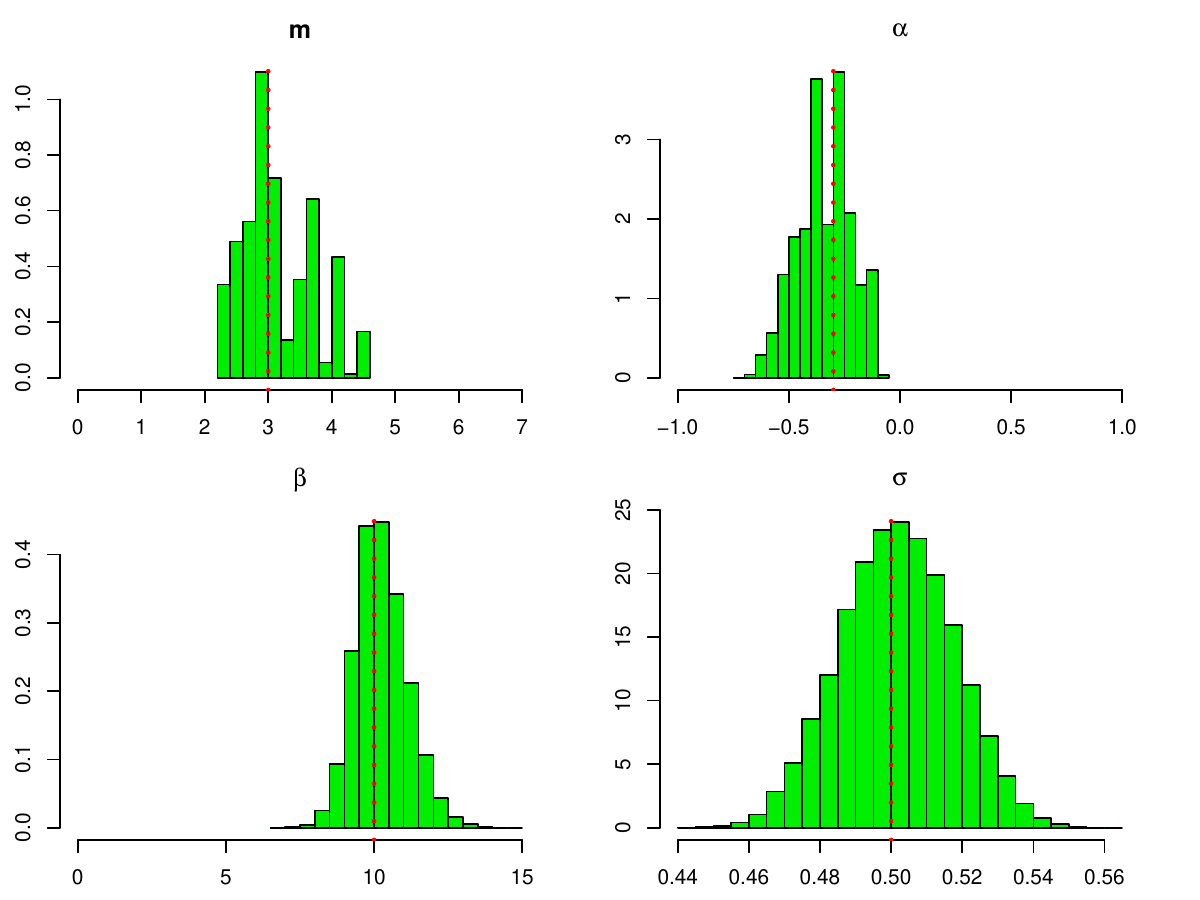}
	\caption{Simulation Study. Histograms depicting subject-specific parameters $m_i$, $\alpha_i$, $\beta_i$, and $\sigma_i$, derived from posterior samples across all subjects. Vertical dotted lines denote the true parameter values.}
	\label{fig:inf_param}
\end{figure}

\begin{figure}[htbp]
	\centering
	\includegraphics[width=.6\linewidth]{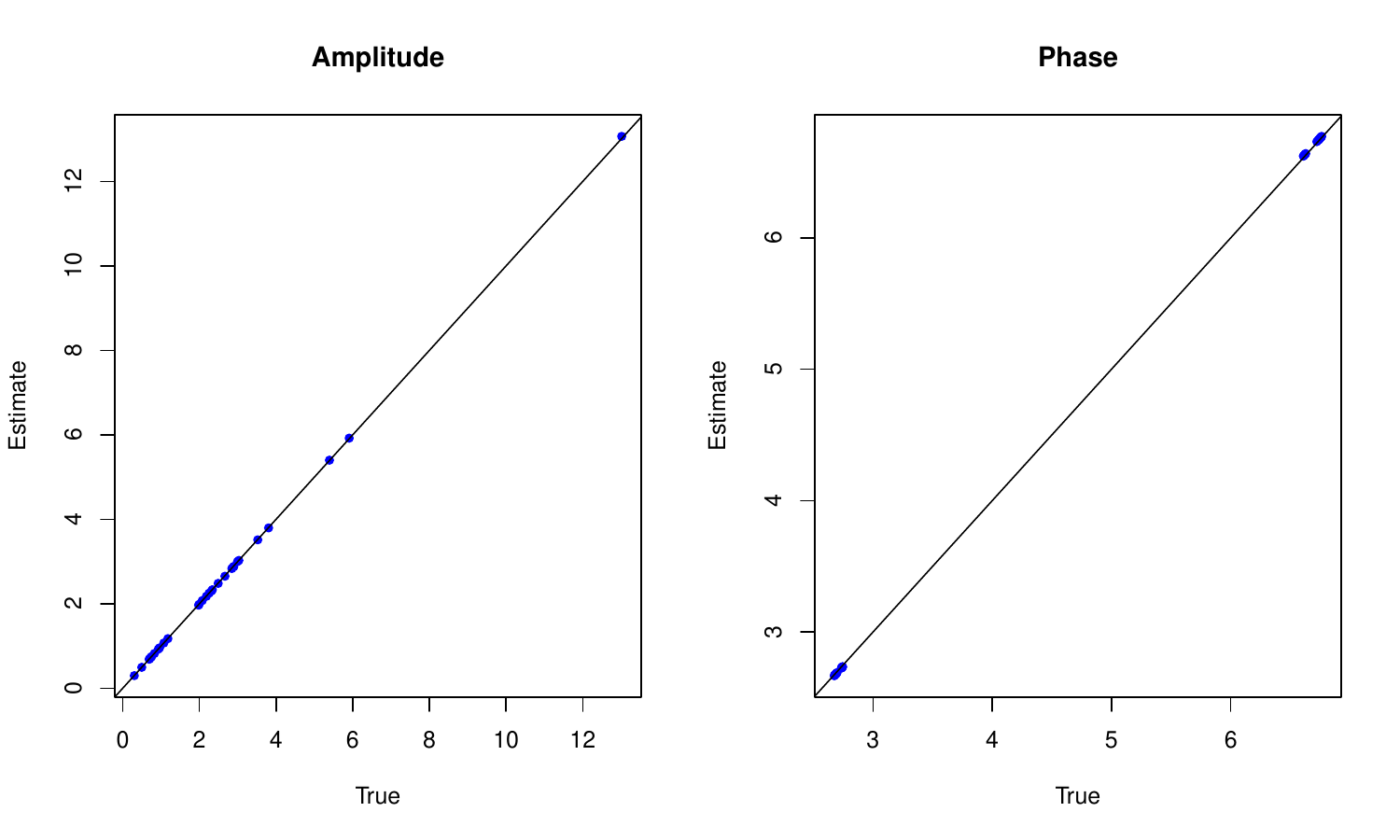}
	\caption{Simulation Study. Scatterplots showcasing estimated values of amplitudes $a(\bm{x_i}, \bm{\eta}a)$ and phases $\phi(\bm{x_i}, \bm{\eta}\phi)$ for $i = 1, \dots, N$, alongside their corresponding true values.}
	\label{fig:scatterplots_ampl_phase}
\end{figure}

Figure \ref{fig:inf_param} illustrates histograms representing subject-specific parameters $m_i$, $\alpha_i$, $\beta_i$, and $\sigma_i$. These histograms are derived from posterior samples across all subjects, with vertical dotted lines indicating the true mean (across subjects) of the parameter values. The mean estimates across subjects are as follows: $\hat{m} = 3.15$, $\hat{\alpha} = -0.33$, $\hat{\beta} = 10.25$, and $\hat{\sigma} = 0.50$. In Figure \ref{fig:scatterplots_ampl_phase}, scatterplots are presented, showcasing the estimated values of amplitudes $a(\bm{x_i}, \bm{\eta}_a)$ and phases $\phi(\bm{x_i}, \bm{\eta}_\phi)$, for $i = 1, \dots, N$, alongside their corresponding true values. These results demonstrate a satisfactory alignment between the estimated and true values. In the Appendix, we present an additional graphical posterior predictive check, showcasing observations alongside 100 draws from the estimated posterior predictive distribution for four representative individuals. 

\begin{figure}[htbp]
	\centering
	\includegraphics[width=0.9\textwidth]{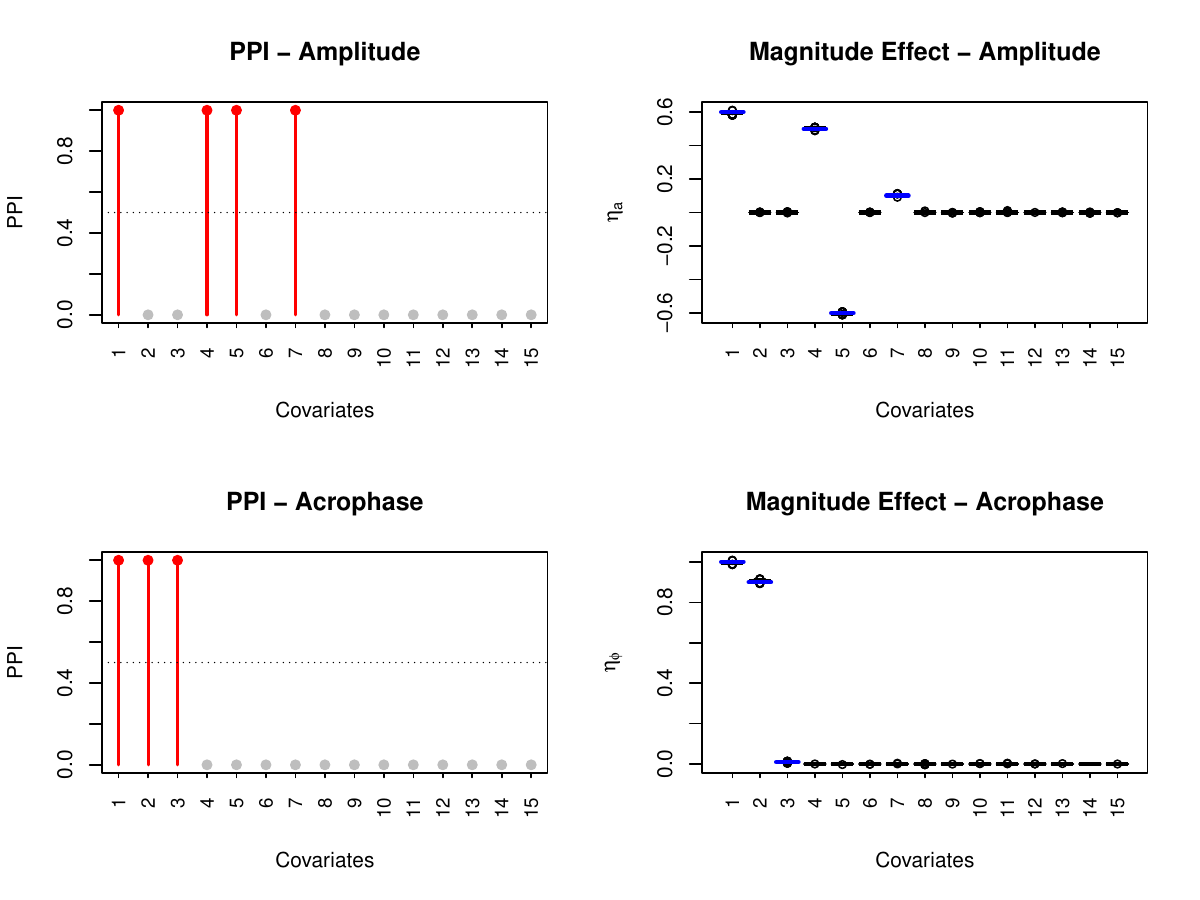}
	\caption{Simulation Study. (left) Posterior probability of inclusion (PPI) for linear coefficients $\bm{\eta}a$ and $\bm{\eta}\phi$;  (right) Boxplots depicting the posterior distribution of effects for amplitude and phase. Horizontal blue lines represent true generating effects for non-zero coefficients. }
	\label{fig:simul_study_PPI_magnitude}
\end{figure}

As for variable selection, in Figure \ref{fig:simul_study_PPI_magnitude} (left), the posterior probabilities of inclusion (PPIs) are displayed for the linear coefficients $\bm{\eta}_a$ and $\bm{\eta}_\phi$, corresponding to amplitude and acrophase, respectively. These were calculated by determining the proportion of MCMC samples in which the corresponding coefficients deviated from zero. Covariate effects were deemed statistically significant if their PPI exceeded 0.5 \citep{barbieri2004optimal}. Notably, our proposed approach adeptly identifies significant covariates, even when effects are small, as exemplified by $\eta_{a, 7}$ and $\eta_{\phi, 3}$. Additionally, in Figure \ref{fig:simul_study_PPI_magnitude} (right), boxplots illustrate the posterior distribution of effects for both amplitude and phase. Horizontal blue lines denote the true generating effect for those effects distinct from zero, showing that the proposed model accurately captures these true generating effects.

\subsubsection{Comparative analysis}

We aim to evaluate the performance of our proposed approach (which we denote CALC $l_1$) against alternative methods, in terms of estimation and variable selection accuracy.  Given the absence in the literature of direct competitors with our proposed model, we opt to conduct a comparative analysis of the proposed approach to the following methods: (i) standard cosinor model \citep{cornelissen2014cosinor} applied independently for each subject; (ii) extended cosinor approach (RAR, \citealt{marler2006sigmoidally}) run independently for each subject; (iii) CALC Blasso,  namely the same model of Eq. \eqref{eq:multisubj_RAR_model} where sparsity is promoted through the lasso prior, with the goal of assessing the efficacy of promoting sparsity through the $l_1$ ball projection  as opposed to the Bayesian Lasso. Note that in the Bayesian setting, the lasso prior is equivalent to placing a double exponential prior on the linear effects \citep{park2008bayesian}. To ensure fairness in the comparison, we set the corresponding hyperparameters of the double exponential distribution equal to those of the $l_1$ ball model, for both amplitude and phase. Models (i) and (ii) were formulated in a Bayesian framework and implemented in \textit{stan}, where the prior distribution for the phase parameter and the parameter governing the duration of rest compared to active time were chosen to be uniform over their respective domains. The rest of the prior distributions were chosen as in our proposed model. 

\begin{figure}[htbp]
	\centering
	\includegraphics[width=1\textwidth]{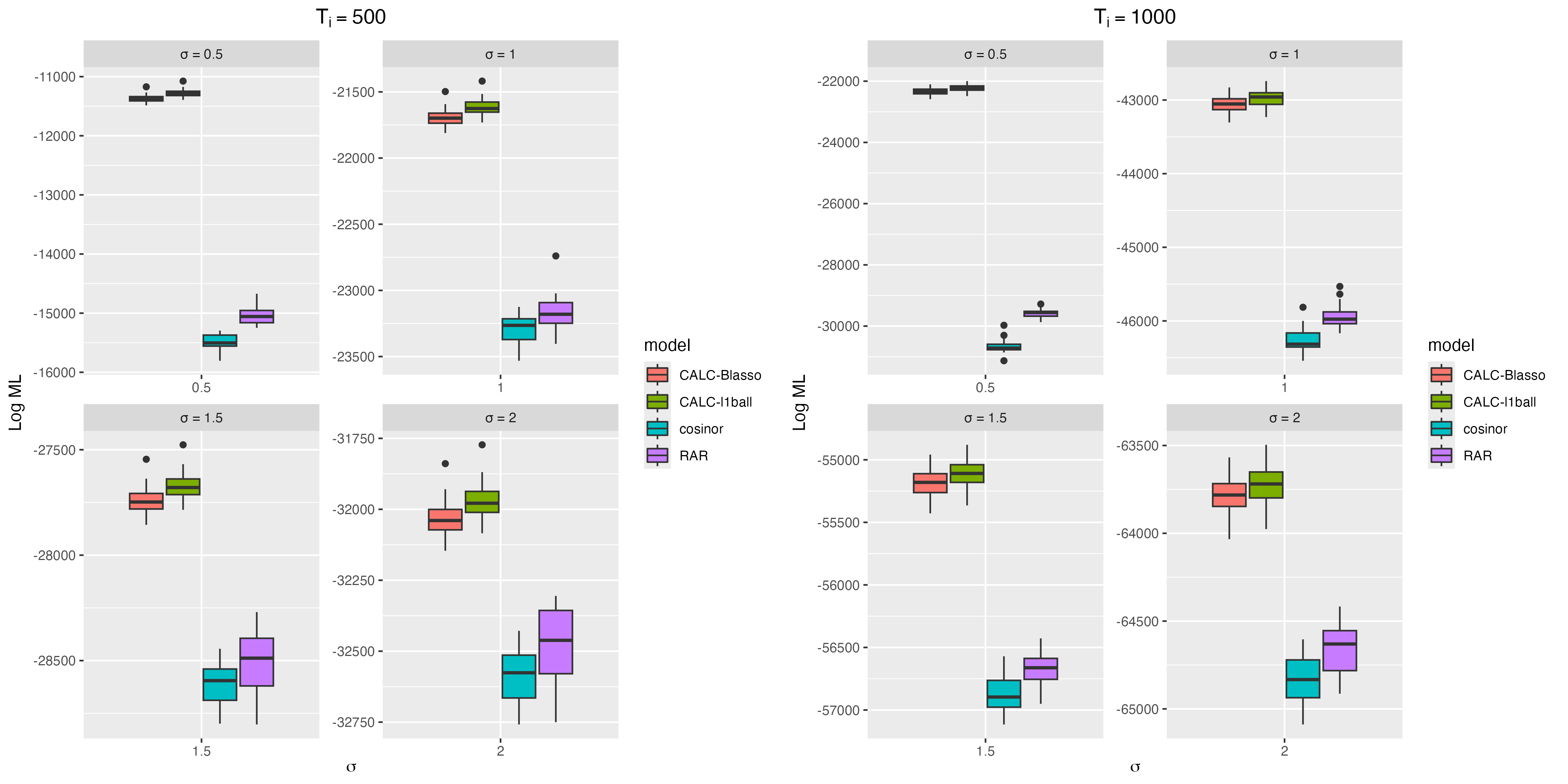}
	\caption{Simulation Study. Marginal likelihood across 20 replicated datasets for $T_i = \{500,  1000\}$ and $\sigma = \{ 0.5, 1, 1.5, 2.0 \}$; results are displayed for CALC $l_1$, CALC Blasso, standard cosinor, and extended cosinor (RAR).  }
	\label{fig:boxplots_logml}
\end{figure}

We consider simulated time series of varying lengths ($T_i = 500$ and $T_i = 1000$) across different levels of residual error ($\sigma = 0.5, 1, 1.5, 2.0$), to explore diverse scenarios of noise influence on the data. First, we investigate the estimation accuracy of the different approaches by computing the marginal likelihood $ p (\bm{y}) := \int \mathcal{L} (\bm{y} | \bm{\theta}) p (\bm{\theta}) d\bm{\theta}$. This is achieved through the application of bridge sampling \citep{meng2002warp}, utilizing both the posterior samples and the seamless integration of bridge sampling with \textit{stan} \citep{gronau2017bridgesampling}. Since the standard cosinor and extended cosinor models are designed for single-subject analysis, we sum the individual log marginal likelihoods to obtain an estimate for the entire cohort. Figure \ref{fig:boxplots_logml} displays boxplots representing the marginal likelihood across 20 replicated datasets for various scenarios, comparing the different approaches. The superiority of the \( l_1 \) ball projection prior over the other approaches is evident across all scenarios, with the distinction becoming less pronounced as \( T_i \) increases.

To assess variable selection performance, we compute key metrics including accuracy, precision, Matthews correlation coefficient (MCC), and F1 score, namely
\begin{equation*}
	\footnotesize
	\begin{split}
		\text{Acc} &= \dfrac{TP + TN}{TP + TN + FP + FN} \\
		\text{Prec} &= \dfrac{TN}{TN + FP} \\
		\text{F1} &= \dfrac{2TP}{2TP + FP + FN} \\
		\text{MCC} &= \dfrac{TP\cdot TN - FP\cdot FN }{\sqrt{(TP+FP) \cdot (TP+FN) \cdot (TN + FP) \cdot (TN + FN)}},
	\end{split}
\end{equation*}

\noindent where $TP$, $FP$, $TN$, $FN$ denote true positive, false positive, true negative and false negative counts. Given the nature of the modeling framework, we can perform variable selection only for CALC $l_1$ and CALC Blasso, while the cosinor and extended cosinor single-subject models are not suitable for this purpose.  For variable selection in the Bayesian Lasso approach, we construct 95\% credible intervals for each effect and consider them significant if the interval does not contain zero.  Additionally, to assess the estimation accuracy of the amplitude and phase coefficients, $\bm{\eta}_a$ and $\bm{\eta}_\phi$, we compute the residual mean squared error (RMSE) and the residual mean absolute error (RMAE) as
	\[
	\begin{split}
		RMSE_z &= \frac{1}{Q}\sum_{i=1}^Q \left(\eta_{z,i} - \hat{\eta}_{z,i}\right)^2, \\
		RMAE_z &= \frac{1}{Q}\sum_{i=1}^Q \left|\eta_{z,i} - \hat{\eta}_{z,i}\right|,
	\end{split}
	\]
	for $z \in \{a,\phi\}$. Moreover, we evaluate the Bayesian coverage of our estimates, defined as the proportion of cases in which the 95\% credible intervals contain the true parameter values. This metric provides a measure of the reliability of our posterior uncertainty quantification in the context of our model.

	% Table \ref{table:comparison} \benni{review comments here} shows the results of this investigation for both amplitude and phase parameters, averaged over 20 replicated datasets. In terms of accuracy, the $l_1$-ball model consistently demonstrates high performance, achieving values ranging from 0.95 to 1.0 for the amplitude, with $T_i = 500$, and maintaining similarly high accuracy scores between 0.93 and 1.0 for $T_i = 1000$. Despite also exhibiting respectable accuracy, the Bayesian Lasso model performs slightly worse than the $l_1$-ball model in achieving consistently high accuracy scores. Furthermore, the l1 ball model consistently maintains strong precision values, ranging from 0.93 to 1.0 across different parameter settings. This indicates the model's ability to make precise positive predictions with minimal false positives. Finally, examination of the MCC values demonstrates the $l_1$-ball model's consistent ability to capture strong correlations between predicted and true classifications, with MCC values ranging from 0.78 to 1.0. While the Bayesian Lasso model achieves respectable MCC values, it is generally outperformed by the l1 ball model. As noise levels increase, the model performance naturally deteriorates, for both the $l_1$-ball and the Bayesian Lasso models. Furthermore, it is notable that both models tend to perform better in identifying amplitude effects compared to phase effects. Overall, these results highlight the $l_1$-ball model's superiority in accurately classifying amplitude and phase effects.  
	
	Table \ref{table:comparison} presents the updated simulation results for both amplitude and acrophase parameters, averaged over 20 replicated datasets. In terms of classification performance, the CALC $l_1$-ball model consistently outperforms CALC Blasso. For amplitude effects, the $l_1$-ball model achieves accuracy values between 0.95 and 1.0 with $T_i = 500$ and between 0.93 and 1.0 with $T_i = 1000$, along with high precision (0.93 to 1.0) and robust MCC values (0.78 to 1.0). As noise levels increase, performance declines for both models, and both tend to perform better for amplitude effects than for acrophase effects. In addition to these classification metrics, the table reports estimation metrics for the amplitude and acrophase coefficients. The RMSE and RMAE quantify estimation accuracy, with lower values indicating more precise recovery of the true parameters. The CALC $l_1$-ball model exhibits substantially lower RMSE and RMAE values compared to CALC Blasso. Moreover, Bayesian coverage confirms the reliability of our posterior uncertainty quantification. Overall, the consistently high coverage achieved by the CALC $l_1$-ball model further supports its superiority in both variable selection and parameter estimation.

\begin{table}[htbp]
	\centering
	\resizebox{\textwidth}{!}{%
		\begin{tabular}{llccccllllcccc}
			\hline
			&            &                      & \multicolumn{2}{c}{Amplitude, $T_i = 500$} &                      &  &  &                      &                      &                      & \multicolumn{2}{c}{Amplitude, $T_i = 1000$} &                      \\ \cmidrule{4-5} \cmidrule{12-13}
			&            & $\sigma=0.5$         & $\sigma=1.0$         & $\sigma = 1.5$       & $\sigma =2.0 $       &  &  &                      &                      & $\sigma=0.5$         & $\sigma=1.0$          & $\sigma = 1.5$       & $\sigma =2.0 $       \\ \cmidrule{2-6} \cmidrule{10-14}
			Acc  & CALC $l_1$-ball    & 1.0                  & 0.958                & 0.996                & 1.0                  &  &  & Acc                  & CALC $l_1$-ball              & 1.0                  & 0.996                & 0.996                & 1.0                  \\
			& CALC Blasso & 0.954                & 0.954                & 0.954                & 0.954                &  &  &                      & CALC Blasso           & 0.972                & 0.972                & 0.972                & 0.972                \\
			Prec & CALC $l_1$-ball    & 1.0                  & 0.947                & 1.0                  & 1.0                  &  &  & Prec                 & CALC $l_1$-ball              & 1.0                  & 1.0                  & 1.0                  & 1.0                  \\
			& CALC Blasso & 1.0                  & 1.0                  & 1.0                  & 1.0                  &  &  &                      & CALC Blasso           & 1.0                  & 1.0                  & 1.0                  & 1.0                  \\
			MCC  & CALC $l_1$-ball    & 1.0                  & 0.924                & 0.992                & 1.0                  &  &  & MCC                  & CALC $l_1$-ball              & 1.0                  & 0.992                & 0.992                & 1.0                  \\
			& CALC Blasso & 0.917                & 0.917                & 0.917                & 0.917                &  &  &                      & CALC Blasso           & 0.946                & 0.946                & 0.946                & 0.946                \\
			F1   & CALC $l_1$-ball    & 1.0                  & 1.0                  & 0.997                & 1.0                  &  &  & F1                   & CALC $l_1$-ball              & 1.0                  & 0.997                & 0.997                & 1.0                  \\
			& CALC Blasso & 0.964                & 0.964                & 0.964                & 0.964                &  &  &                      & CALC Blasso           & 0.979                & 0.979                & 0.979                & 0.979                \\
			RMSE & CALC $l_1$-ball    & 3.42e-06            & 4.46e-03            & 3.82e-05            & 6.88e-05            &  &  & RMSE                 & CALC $l_1$-ball              & 3.27e-06            & 1.27e-05            & 2.85e-05            & 4.84e-05            \\
			& CALC Blasso & 1.67e-04            & 1.67e-04            & 1.67e-04            & 1.67e-04            &  &  &                      & CALC Blasso           & 7.19e-05            & 7.19e-05            & 7.19e-05            & 7.19e-05            \\
			RMAE  & CALC $l_1$-ball    & 7.98e-04            & 1.45e-02            & 2.86e-03            & 3.76e-03            &  &  & RMAE                  & CALC $l_1$-ball              & 7.78e-04            & 1.53e-03            & 2.33e-03            & 3.12e-03            \\
			& CALC Blasso & 9.02e-03            & 9.02e-03            & 9.02e-03            & 9.02e-03            &  &  &                      & CALC Blasso           & 6.32e-03            & 6.32e-03            & 6.32e-03            & 6.32e-03            \\
			Coverage & CALC $l_1$-ball & 1.0                 & 0.940                & 0.996                & 0.996                &  &  & Coverage             & CALC $l_1$-ball              & 0.993                & 0.996                & 0.996                & 0.996                \\
			& CALC Blasso      & 0.940               & 0.940                & 0.940                & 0.940                &  &  &                      & CALC Blasso           & 0.951                & 0.951                & 0.951                & 0.951                \\ \cmidrule{1-14}
			&            & \multicolumn{1}{l}{} & \multicolumn{1}{l}{} & \multicolumn{1}{l}{} & \multicolumn{1}{l}{} &  &  &                      &                      & \multicolumn{1}{l}{} & \multicolumn{1}{l}{}  & \multicolumn{1}{l}{} & \multicolumn{1}{l}{} \\
			&            &                      & \multicolumn{2}{c}{Acrophase, $T_i = 500$}     &                      &  &  &                      &                      &                      & \multicolumn{2}{c}{Acrophase,   $T_i = 1000$}    &                      \\ \cmidrule{4-5} \cmidrule{12-13}
			&            & $\sigma=0.5$         & $\sigma=1.0$         & $\sigma = 1.5$       & $\sigma =2.0 $       &  &  & \multicolumn{1}{c}{} & \multicolumn{1}{c}{} & $\sigma=0.5$         & $\sigma=1.0$          & $\sigma = 1.5$       & $\sigma =2.0 $       \\ \cmidrule{2-6} \cmidrule{10-14}
			Acc  & CALC $l_1$-ball    & 0.996               & 0.937                & 0.937                & 0.930                &  &  & Acc                  & CALC $l_1$-ball              & 1.0                  & 0.965                & 0.940                & 0.933                \\
			& CALC Blasso & 0.937                & 0.937                & 0.937                & 0.937                &  &  &                      & CALC Blasso           & 0.951                & 0.951                & 0.951                & 0.951                \\
			Prec & CALC $l_1$-ball    & 0.996               & 0.938                & 0.931                & 0.926                &  &  & Prec                 & CALC $l_1$-ball              & 1.0                  & 0.971                & 0.939                & 0.931                \\
			& CALC Blasso & 0.991                & 0.991                & 0.991                & 0.991                &  &  &                      & CALC Blasso           & 1.0                  & 1.0                  & 1.0                  & 1.0                  \\
			MCC  & CALC $l_1$-ball    & 0.989               & 0.802                & 0.797                & 0.775                &  &  & MCC                  & CALC $l_1$-ball              & 1.0                  & 0.890                & 0.810                & 0.788                \\
			& CALC Blasso & 0.860                & 0.860                & 0.860                & 0.860                &  &  &                      & CALC Blasso           & 0.902                & 0.902                & 0.902                & 0.902                \\
			F1   & CALC $l_1$-ball    & 0.998               & 0.961                & 0.962                & 0.958                &  &  & F1                   & CALC $l_1$-ball              & 1.0                  & 0.978                & 0.964                & 0.960                \\
			& CALC Blasso & 0.956                & 0.956                & 0.956                & 0.956                &  &  &                      & CALC Blasso           & 0.964                & 0.964                & 0.964                & 0.964                \\ 
			RMSE & CALC $l_1$-ball    & 4.16e-06            & 2.35e-05            & 3.90e-05            & 6.19e-05            &  &  & RMSE                 & CALC $l_1$-ball              & 1.59e-06            & 9.81e-06            & 2.30e-05            & 3.74e-05            \\
			& CALC Blasso & 8.67e-05            & 8.67e-05            & 8.67e-05            & 8.67e-05            &  &  &                      & CALC Blasso           & 3.21e-05            & 3.21e-05            & 3.21e-05            & 3.21e-05            \\
			RMAE  & CALC $l_1$-ball    & 6.92e-04            & 1.95e-03            & 2.43e-03            & 2.95e-03            &  &  & RMAE                  & CALC $l_1$-ball              & 4.32e-04            & 1.19e-03            & 1.92e-03            & 2.42e-03            \\
			& CALC Blasso & 6.23e-03            & 6.23e-03            & 6.23e-03            & 6.23e-03            &  &  &                      & CALC Blasso           & 3.92e-03            & 3.92e-03            & 3.92e-03            & 3.92e-03            \\
			Coverage & CALC $l_1$-ball & 0.993               & 0.958                & 0.947                & 0.937                &  &  & Coverage             & CALC $l_1$-ball              & 0.989                & 0.986                & 0.965                & 0.937                \\
			& CALC Blasso      & 0.926               & 0.926                & 0.926                & 0.926                &  &  &                      & CALC Blasso           & 0.951                & 0.951                & 0.951                & 0.951                \\ \hline
	\end{tabular}}
	\caption{Simulation Study. Classification metrics across 20 simulations for $T_i = \{500,  1000\}$ and $\sigma = \{ 0.5, 1, 1.5, 2.0 \}$. We report accuracy, precision, MCC, and F1 score, RMSE, RMAE, and coverage; results are displayed for CALC $l_1$-ball and the Bayesian Lasso (CALC Blasso). }
	\label{table:comparison}
\end{table}

 In the Supplementary Material, we assess out‐of‐sample predictive performance by comparing our CALC \(l_1\) model with CALC Blasso, the standard cosinor model, and an extended cosinor approach (RAR). For sample sizes \(T_i = 500\) and \(T_i = 1000\), and across noise levels \(\sigma \in \{0.5, 1.0, 1.5, 2.0\}\), models were trained on the observed data and evaluated on an additional 25\% simulated test set. Using log-transformed RMSE and RMAE as performance metrics, our CALC \(l_1\) model consistently achieved lower error values than the alternatives, indicating superior predictive accuracy.

\section{Application to Actigraphy Data from an Epilepsy Study}
\label{sec:application}

 Wrist actigraphy recordings provide valuable insights into the timing and quantity of activity and sleep. Sleep-wake rhythms are an important component of health, and modulate (or are modulated by) developmental processes, aging and health \citep{wallace2023rest}.  To test the utility of our predictive model, here, we apply our approach to actigraphy recordings from a collection of adults with epilepsy. Epilepsy, or the epilepsies, are an etiologically heterogeneous collection of brain disorders that are defined by an enduring risk to develop epileptic seizures. People with epilepsy are also prone to a host of neuropsychiatric and somatic comorbidities that may or may not neatly parallel the burden of epileptic seizures. Not surprisingly, periods of high seizure frequency are often associated with disruptions in activity and sleep rhythms. Orally ingested antiseizure medications (ASMs), which constitute the backbone of seizure prevention strategies, can also impact mood, anxiety and sleepiness, especially when multiple ASMs are combined. In published cross-sectional analyses comparing against recordings from control subjects, people with epilepsy display RARs with lower amplitudes and poorer robustness  \citep{liguori2022more, abboud2023actigraphic, tang2024circadian}. Here, we take advantage of one of these datasets that is richly annotated with a set of 14 covariates.

% In the context of epilepsy research, actigraphy offers a non-invasive method to monitor rest-activity rhythms, enabling the identification of patterns that may correlate with dynamic fluctuations in seizure frequency and/or the burden of psychiatric comorbidities, like depression or anxiety.  
%In this section, we analyze the dataset previously presented in \citet{abboud2023actigraphic}. We outline the study layout, data collection methods, preprocessing steps, and the results obtained from applying our proposed approach.

\subsection{Study layout and data preprocessing}
We analyze the data previously presented in \citet{abboud2023actigraphic}. Participants were recruited and provided consent at the Baylor Comprehensive Epilepsy Clinic in Houston, TX. Eligible participants were adults between the ages of 18 and 75 who experienced partial-onset seizures. Participants were asked to continuously wear the robust and widely-utilized FDA-approved Actiwatch-2 device (Philips Respironics) on their non-dominant wrist for at least 10 days, including during sleep and showers. The Actiwatch-2 device employs piezoelectric sensors to capture unidimensional accelerometry, with the sensor specifically aligned to monitor natural wrist movements.  Here, we examined activity time series data from the subset of subjects that provided at least 6 full days of recorded activity, resulting in a total of $N=47$ individuals.

In addition to wearing the watch, participants were requested to complete a series of psychometric assessments (on paper forms), including the QIDS-SR (Quick Inventory of Depression Symptoms-Self-Report), ESS (Epworth Sleepiness Scale), AEP (Adverse Event Profile), and PHQ-SADS, which encompasses the PHQ-15 for somatic symptoms, GAD-7 for anxiety, and PHQ-9 for depression \citep{gilliam2004systematic, kocalevent2013standardization, kendzerska2014evaluation, plummer2016screening, matsumoto2024assessing}. In all six surveys, higher scores are directly proportional to more intense and pervasive symptoms. Table \ref{table:demographics} provides an overview of the demographic and clinical data, comprising 14 covariates: Sex, ESS, PHQ-15, PHQ- 9, GAD-7, QIDS-SR, AEP, age, number of non-rescue antiseizure medications, employment Status, body mass index, a history of epilepsy surgery, current use of SSRIs (selective serotonin reuptake inhibitors), and drug refractoriness (judged at the time of data collection).  In our modeling approach, we also included an intercept term, so that the total number of covariates is $Q$ = 15.

\begin{table}[htbp]
	\small
	\centering
	\begin{tabular}{ll}
		\hspace{-0.2cm}\\
		\hline
		\textbf{Characterstics} & \textbf{Value}   \\   \hline
		%Sample Size             & 47               \\
		%Days of Recording       & 8  (6-14)        \\
		Age                     & 43 (18-72)       \\
		Female Sex                & 31 (0.66)        \\
		BMI(kg/m$^2$)              & 29.9 (16.0-53.0) \\
		Employed                & 17 (0.36)        \\
		ESS                     & 6.8 (0-19)       \\
		PHQ15                   & 7.7 (0-21)       \\
		PHQ9                    & 6.7 (0-27)       \\
		GAD7                    & 5.1 (0-19)       \\
		QIDSSR                  & 10.6 (0-26)      \\
		AEP                     & 41.3 (19-63)     \\
		N. Medications          & 2 (1-5)          \\
		Prior Epilepsy Surgery  & 17 (0.36)        \\
		Taking an SSRI               & 11 (0.23)        \\
		Medically refractory              & 32 (0.68)  \\ \hline       
	\end{tabular}
	\caption{
		\textbf{Demographic and clinical data summary}. For non-binary variables, the mean (range) is displayed. For binary variables, the count of individuals (proportion) is provided.}
	\label{table:demographics}
\end{table}

%\subsection{Data pre-processing}

\begin{figure}[htbp]
	\centering
	\includegraphics[width=1\textwidth]{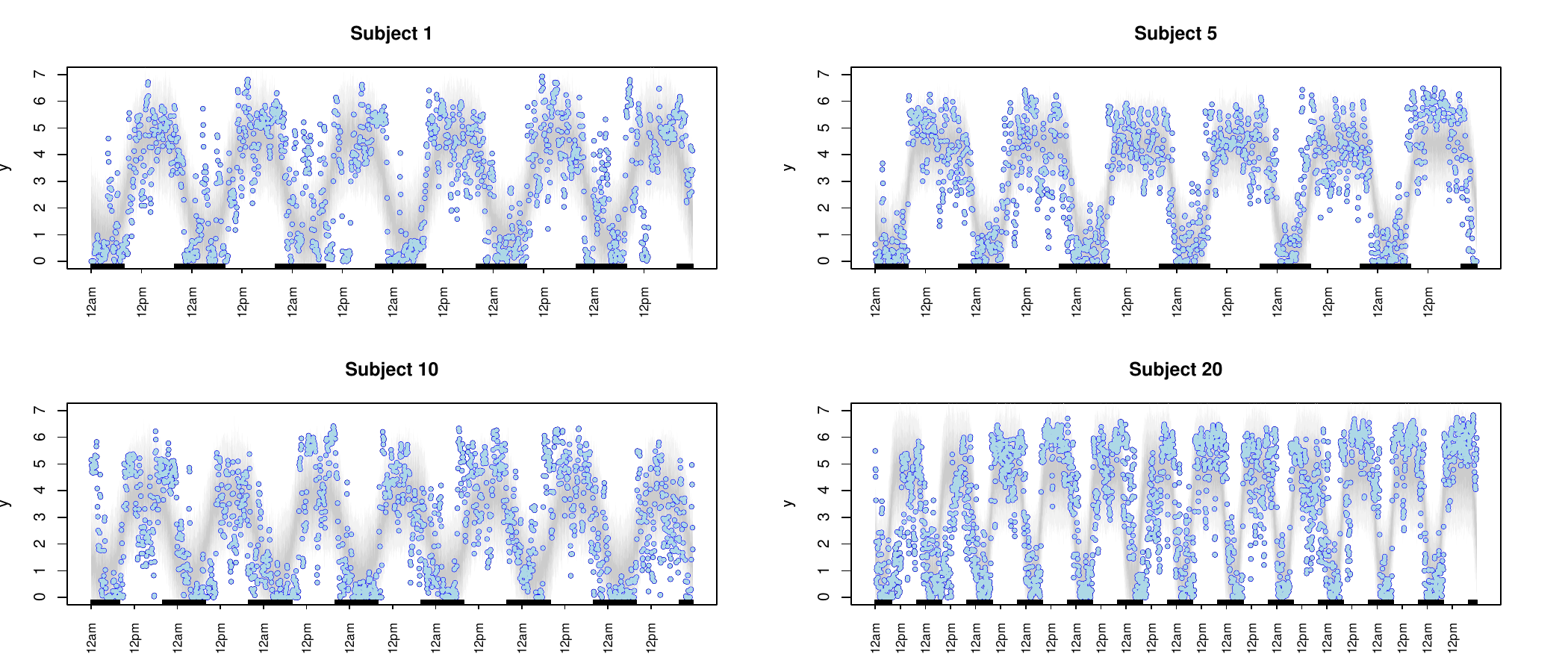}
	\caption{
		\textbf{Actigraphy Data.} Log activity time series for four representative subjects, illustrated by dots. Shaded gray lines depict 100 draws from the posterior predictive distribution. Rectangles along the time axis indicate periods from 8pm to 8am. }
	\label{fig:data_post_pred}
\end{figure}

We applied the following preprocessing steps to the original activity data, following a similar routine to previous work \citep{huang2018hidden, abboud2023actigraphic}:  we (i) applied a logarithmic transformation, adding one to each value, to reduce the high variability in counts, particularly between periods of activity and near-zero sleep states; (ii) applied a moving average filter with a span of 15 points (equivalent to fifteen minutes), to reduce noise and improve the signal; (iii) downsampled the data by averaging every 5 points, resulting in a sampling rate of 5 minutes, thus improving computational efficiency; (iv) and standardized the data using the mean and standard deviation calculated from the combined data of all subjects. Two missing values in covariate PHQ9 were imputed utilizing the \texttt{missforest} package which employs a random forest algorithm. Figure \ref{fig:data_post_pred} displays the pre-processed time series for four representative individuals. We also standardized all non-binary covariates to have a mean of zero and a variance of one.

\subsection{Parameter settings} 
We set $\xi_a = \xi_\phi = 0.75$ to reflect a moderate degree of sparsity of the effects for both amplitude and phase. As discussed in Section \ref{sec:hyperparms_elictation}, since we have reasonable prior information about amplitude and phase, we propose to integrate prior information about the magnitude of amplitude and phase in a more informative manner into the model. In particular, we propose to formulate the hyperprior specification in a way that ensures the prior probability of the average magnitude (and phase) not exceeding $q_a$ (and $q_\phi$) is governed by a predetermined probability $p_a$ (and $p_\phi$). The previous study of \cite{abboud2023actigraphic} has suggested that the amplitude of individuals does not exceed $q_a = 8$ and the phase does not exceed $q_\phi = 20$ (noting also that all the time series begin at midnight); we express a confidence level in these assertions of $p_a = p_\phi = 0.95$. Implementation of Algorithm \ref{Alg:hyperparms_elictation} yielded the following selection of hyperparameters for the $l_1$-ball: $\tau_a = 0.39$, $\tau_\phi = 0.5$, $\lambda_a = 0.93$, and $\lambda_\phi = 0.76$. The rest of the hyperparameters were set to be weakly informative as described in the simulation studies. The proposed model was run with \text{stan} to sample 4000 MCMC samples, discarding 2000 updates as burn-in period. Convergence of the MCMC chains was assessed via visual inspection of the trace plots as well as Gelman-Rubin convergence diagnostics $\hat{R}$.

\subsection{Results}

% Figure \ref{fig:data_post_pred} illustrates the model fitting for the selected four as a graphical check consisting of 100 draws from the posterior predictive distribution, demonstrating that the model is effectively capturing the  dynamic circadian fluctuations. 

Figure \ref{fig:data_post_pred}  illustrates the model fitting for four selected individuals, with 100 draws from the posterior predictive distribution demonstrating the model’s ability to capture dynamic circadian fluctuations. Although the Gaussian assumption on the log scale can occasionally produce negative predictions, a simple practical measure is to truncate these values at zero if strictly non-negative predictions are desired.

\begin{figure}[htbp]
	\centering
	\includegraphics[width=0.8\textwidth]{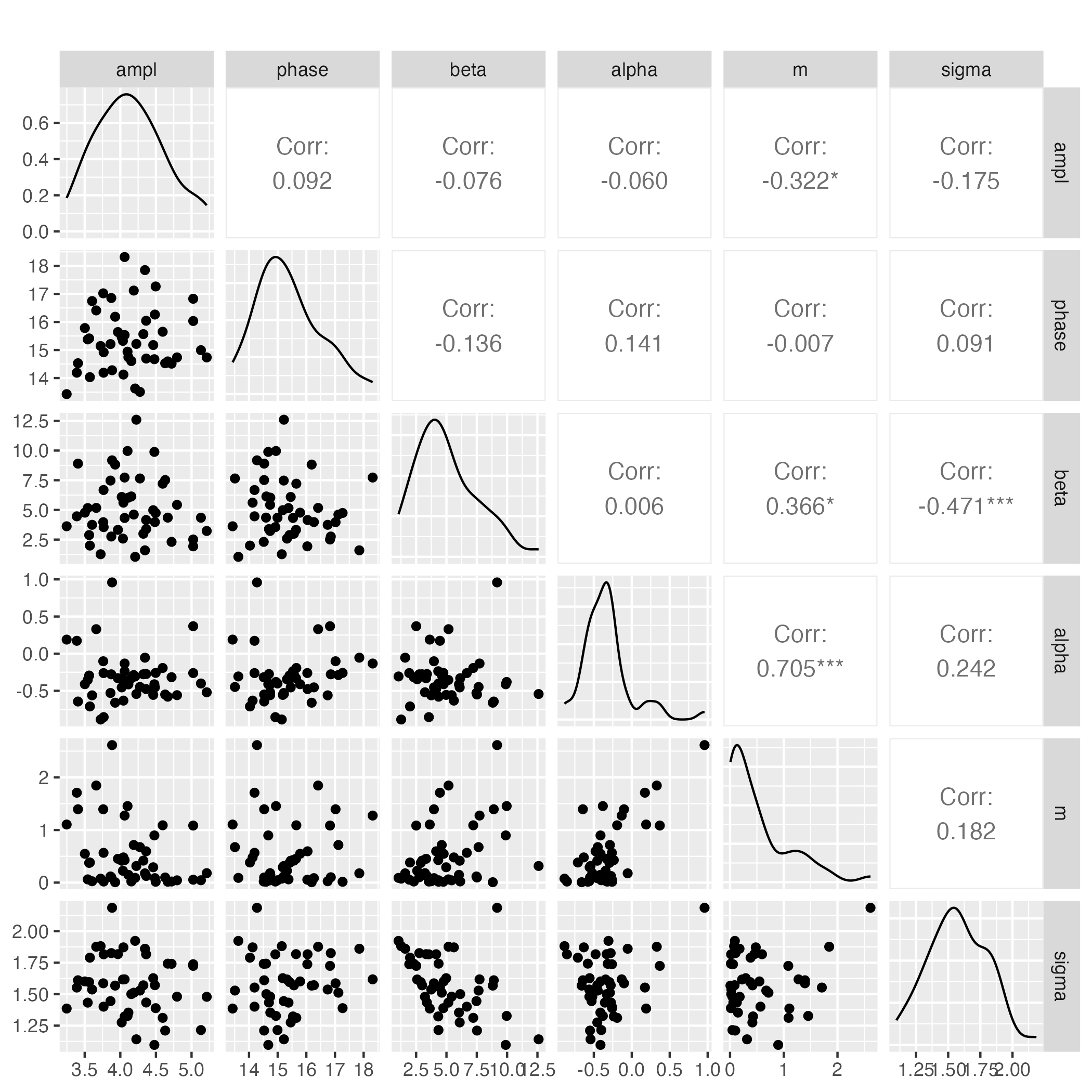}
	\caption{\textbf{Actigraphy Data.} Scatterplots showing the relationships between each combination of parameters $a(\bm{X}, \bm{\eta}_a)$, $\phi(\bm{X}, \bm{\eta}_\phi)$, $\bm{\beta}$, $\bm{\alpha}$, $\bm{m}$, and $\bm{\sigma}$. Histograms of the posterior samples are reported in the plots on the diagonal. Pairwise Pearson correlation coefficients are also reported. }
	\label{fig:scatterplot_estimate}
\end{figure}

Figure \ref{fig:scatterplot_estimate} presents scatterplots illustrating the relationships between the parameter combinations \(a(\bm{X}, \bm{\eta}_a)\), \(\phi(\bm{X}, \bm{\eta}_\phi)\), \(\bm{\beta}\), \(\bm{\alpha}\), \(\bm{m}\), and \(\bm{\sigma}\). Histograms of posterior samples across all subjects are displayed along the diagonal plots. The mean parameter estimates across subjects are as follows: \(\bar{a}(\bm{X}, \bm{\eta}_a) = 4.14\), \(\bar{\phi}(\bm{X}, \bm{\eta}_\phi) = 15.36\), \(\bar{\bm{\beta}} = 5.03\), \(\bar{\bm{\alpha}} = -0.33\), \(\bar{\bm{m}} = 0.52\), and \(\bar{\bm{\sigma}} = 1.57\). Pairwise Pearson correlation coefficients were also computed to evaluate the strength and direction of the relationships among the parameter estimates.  This graphical exploration reveals several notable associations among the parameters governing rest-activity rhythms. A significant negative correlation (\(-0.322\)) was observed between the amplitude \(a(\bm{X}, \bm{\eta}_a)\) and the minimum expected activity level \(\bm{m}\), suggesting that individuals with greater oscillations in activity tend to exhibit lower baseline activity levels.  Conversely, the shape parameter \(\bm{\beta}\), which reflects the rate of change during transitions between rest and activity, showed a meaningful positive correlation (\(0.366\)) with \(\bm{m}\). This indicates that individuals with steeper and more abrupt transitions between rest and activity (higher \(\bm{\beta}\)) tend to exhibit higher baseline activity levels.  The parameter \(\bm{\sigma}\), representing variability in activity patterns, exhibited a strong negative correlation (\(-0.471\)) with \(\bm{m}\), indicating that individuals with more irregular activity patterns tend to have lower baseline activity levels. In contrast, the duration parameter \(\bm{\alpha}\), which governs the proportion of time spent in rest versus activity, displayed a strong positive correlation (\(0.705\)) with \(\bm{m}\). This suggests that individuals with higher \(\bm{\alpha}\), indicating shorter active periods and longer rest periods, tend to maintain higher baseline activity levels, potentially reflecting distinct patterns of rest-activity balance.  These findings highlight the complex interplay among the parameters shaping circadian dynamics, offering valuable insights into individual variability in rest-activity rhythms and their underlying physiological processes.

\subsubsection{Factors Impacting Amplitude and Phase in Activity Profiles}  
\begin{figure}[htbp]
	\centering
	\includegraphics[width=0.8\textwidth]{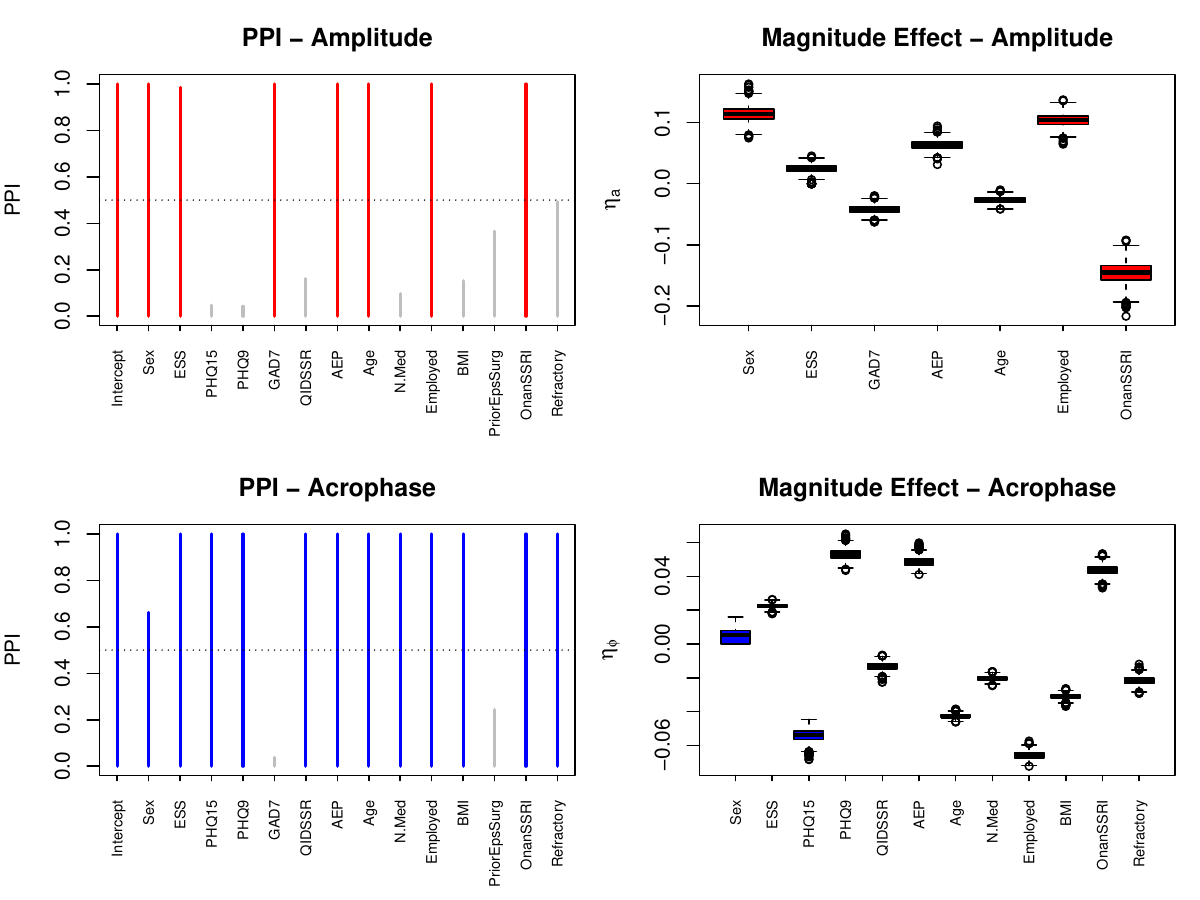}
	\caption{\textbf{Actigraphy Data.} (left) Posterior probabilities of inclusion (PPIs) of the regression coefficients for amplitude, $\bm{\eta}_a$, and phase,  $\bm{\eta}_\phi$, with significant covariates highlighted in red for amplitude and blue for phase; (right) Boxplots of the posterior samples of the corresponding selected coefficients (excluding the intercepts).}
	\label{fig:data_PPI_magnitude}
\end{figure}

Figure \ref{fig:data_PPI_magnitude} (left) shows the posterior probabilities of inclusion (PPIs) of the regression coefficients for amplitude, $\bm{\eta}_a$, and phase,  $\bm{\eta}_\phi$. Here, significant covariates (selected with marginal posterior probabilities of greater or equal to 50\%) are highlighted in red for amplitude and blue for phase. Additionally, Figure \ref{fig:data_PPI_magnitude} (right) presents boxplots illustrating the posterior distribution of the corresponding selected coefficients, excluding the intercepts. 
% \textcolor{blue}{In what follows, we report the posterior median and the 95\% credible interval in square brackets.} %The results depicted in Figure \ref{fig:data_PPI_magnitude} highlight significant associations between covariates and activity data derived from accelerometers within our multi-subject rest-activity model. 
Results indicate that sex positively influences amplitude, with a posterior median effect of 0.11 and a 95\% credible interval equal to [0.09, 0.14], replicating previous findings demonstrating that females display higher measured RAR amplitudes across much of the adult lifespan \citep{li2021demographic}. Conversely, negative effects on amplitude are evident for individuals taking an SSRI (-0.15, [-0.11, -0.18]) and those experiencing a higher burden of anxiety symptoms measured by the GAD7 (-0.04, [-0.03, -0.06]). Regarding phase, significant positive effects (i.e., phase delay) is linked to higher PHQ9 (0.053, [0.048, 0.060]) and AEP scores (0.048, [0.044, 0.054]), confirming the known association between delayed acrophase and mood/somatic symptoms \citep{smagula2022association}. Conversely, negative effects on phase are observed for employment status (-0.066, [-0.070, -0.062]) and PHQ15 (-0.053, [-0.061, -0.047]).

% As a further validation of the results discussed above, we took a closer look at the distributions of the subject-specific estimates of the covariate dependent parameters, i.e. amplitudes $ a(\bm{X}, \eta_a) = \big( a(\bm{x}_1, \bm{\eta}_a), \dots, a(\bm{x}_N, \eta_a)\big)$ and phases $\phi(\bm{X}, \eta_\phi) = \big( \phi(\bm{x}_1, \bm{\eta}_\phi), \dots, \phi(\bm{x}_N, \eta_\phi)\big)$, for the selected covariates. Figures \ref{fig:ampl_scatter_covariate} and  \ref{fig:phase_scatter_covariate}  illustrate scatterplots of the estimated effects on amplitude and phase, respectively, for the significant covariates selected with marginal PPI greater or equal to 50\%. 

\begin{figure}[htbp]
	\centering
	\includegraphics[width=0.8\textwidth]{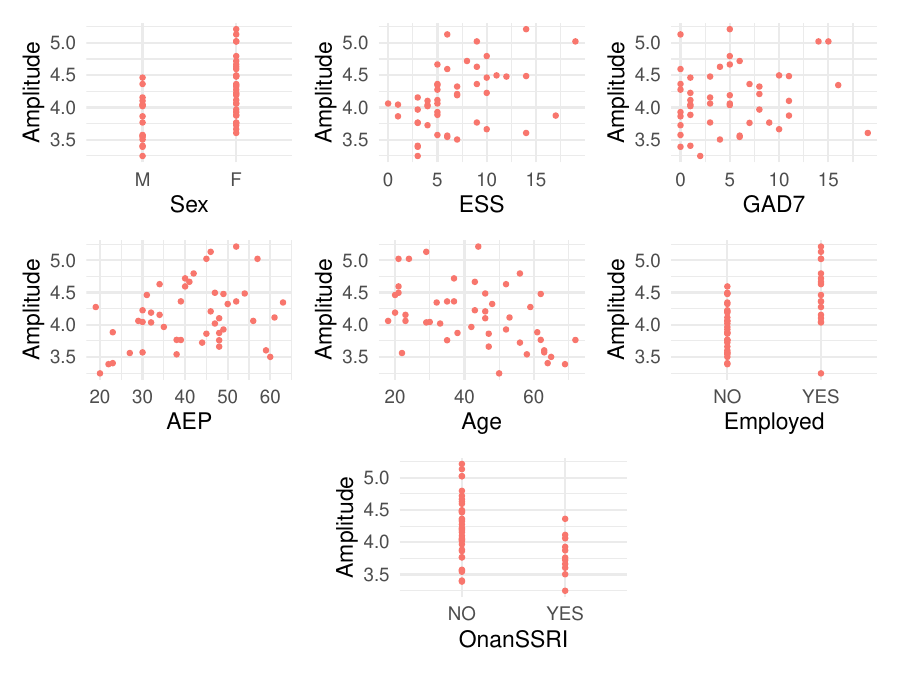}
	\caption{\textbf{Actigraphy Data.} Scatterplots of subject-specific estimated effects on amplitudes, by significant covariates. }
	\label{fig:ampl_scatter_covariate}
\end{figure}

\begin{figure}[htbp]
	\centering
	\includegraphics[width=0.8\textwidth]{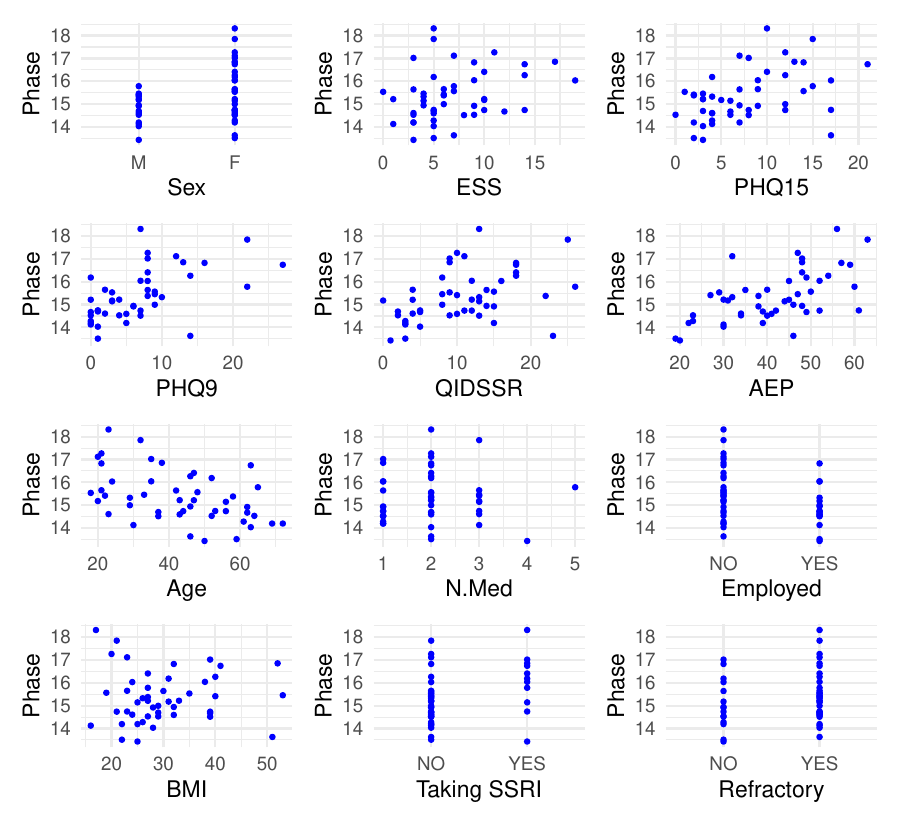}
	\caption{\textbf{Actigraphy Data.} Scatterplots of subject-specific estimated effects on phases, by significant covariates. }
	\label{fig:phase_scatter_covariate}
\end{figure}

As a further validation of the results discussed above, we took a closer look at the distributions of the subject-specific estimates of the covariate-dependent parameters, i.e. amplitudes \( a(\bm{X}, \eta_a) = \big( a(\bm{x}_1, \bm{\eta}_a), \dots, a(\bm{x}_N, \eta_a)\big) \) and phases \( \phi(\bm{X}, \eta_\phi) = \big( \phi(\bm{x}_1, \bm{\eta}_\phi), \dots, \phi(\bm{x}_N, \eta_\phi)\big) \), for the selected covariates. 
Figure \ref{fig:ampl_scatter_covariate} displays scatterplots of the estimated amplitude for each subject (from Equation 2), where the subplots correspond to the division by a certain covariate (either discrete like gender or employment status, or continuous like AEP or GAD7). For example, on the x-axis is gender (male or female), and on the y-axis is the estimate of the amplitude. The subpanels corresponding to the different covariate splits are chosen for the significant covariates selected with marginal PPI greater than or equal to 50\%. Figure \ref{fig:phase_scatter_covariate}  follows the same structure but displays the estimated phase instead of amplitude.
These plots highlight several interesting relationships. For example, coefficient estimates for some of the binary covariates, such as sex, show a noticeable separation.  To quantify the differences in the distributions of these binary covariates, we utilized the MCMC output to calculate averages of the subject-specific amplitudes and phases across the two groups of individuals identified by the value of the binary covariate, and observed whether the 95\% credible interval of the difference in means included zero. We adapted this empirical testing procedure to the non-binary variables by splitting the subjects into two groups, where we used a median split. For amplitude, the mean value for females resulted significantly higher than male amplitude, with a posterior mean of the difference of means equal to 0.121 and credible interval $(0.102, 0.141)$. Furthermore, subjects taking an SSRI exhibit lower amplitude compared to those not assuming it (-0.107; ($-0.084, -0.130$)).  A notable difference is observed in the amplitude of AEP, where subjects with AEP greater than the empirical median value of 42 show significantly higher amplitude than those with AEP less than 42 (-0.043; ($-0.026 , -0.058))$. Regarding phase, individuals with a PHQ15 score between 7 and 21 have significantly higher phase values compared to those with a score below 7 (-0.067; ($-0.064, -0.071))$. Additionally, unemployed individuals generally exhibit higher phase values (-0.049; ($-0.053, -0.046$)), though with greater variability. Lastly, individuals with high AEP (42-63) have significantly higher phase values than those with low AEP (19-42), (-0.071; ($-0.074, -0.068))$. All subject-specific parameter estimates are provided in the Supplementary Material. Overall these findings underscore the intricate  ways in which demographic, psychological, and medical factors may shape activity profiles, offering valuable insights for personalized healthcare interventions and clinical assessments. 

% \begin{figure}[htbp]
%     \centering
%     \includegraphics[width=0.9\textwidth]{figures/boxplots_covariate.pdf}
%     \caption{\textbf{Actigraphy Data.} Boxplots of subject-specific amplitude and phase sub-stratified according to sex, refractory employed, AEP, and PHQ15.}
%     \label{fig:boxplots_covariate}
% \end{figure}

\section{Concluding Remarks} \label{sec:concluding_remarks}
% We have introduced a Bayesian covariate-dependent anti-logistic circadian model for analyzing activity data from continuous accelerometry data.  This innovative approach simultaneously integrates demographic and clinical variables, enhancing our understanding of individual circadian dynamics. Through rigorous simulation studies and application to real-world data, our method has demonstrated robustness and potential for informing personalized healthcare interventions. This advancement underscores the significance of incorporating covariates in modeling rest-activity rhythms.

% Applying the proposed multivariate framework to substantially larger publicly deposited actigraphy datasets may reveal more generalizable population-level relationships \citep{redline2014sleep, zhang2018national}. By simultaneously considering variables like temperature variations and patterns of rest-activity, researchers can account for the complex interplay of multiple factors influencing biological rhythms. Such an approach promises to enhance our ability to model and predict how these intricate systems respond to various environmental and internal cues, thereby advancing our comprehension of circadian and ultradian phenomena. Further avenues for exploration include investigating frequencies beyond the standard 24-hour circadian rhythm, such as ultradian oscillations that occur more frequently within a day. These shorter cycles may play crucial roles in regulating physiological processes and could offer a richer understanding of biological timing mechanisms. 

% \benni{future directions to add:

We have introduced a Bayesian covariate-dependent anti-logistic circadian model for analyzing activity data from continuous accelerometry. This innovative approach integrates demographic and clinical variables, thereby enhancing our understanding of individual circadian dynamics. Through simulation studies and real-world applications, our method demonstrates robustness and has the potential to inform personalized healthcare interventions, underscoring the importance of incorporating covariates in modeling rest-activity rhythms.

Applying the proposed multivariate framework to substantially larger publicly deposited actigraphy datasets may reveal more generalizable population-level relationships \citep{redline2014sleep, zhang2018national}. By simultaneously considering factors such as seasonal variations and rest-activity patterns, researchers can better account for the complex interplay of multiple influences on biological rhythms. Such an approach promises to advance our ability to model and predict how these intricate systems respond to environmental and internal cues, thereby broadening our understanding of both circadian and ultradian phenomena. Further exploration of frequencies beyond the standard 24-hour cycle (e.g., ultradian oscillations) may also yield valuable insights into shorter yet highly influential cycles that regulate various physiological processes. Finally, while we have applied this approach to people with epilepsy, similar tactics could be utilized in other medical conditions where perturbations in rest activity rhythms may have a multifactorial etiology (e.g., multiple sclerosis, chronic kidney disease, systemic lupus erythematosus, etc). 

 Although our proposed framework demonstrates promising results for modeling circadian activity data with covariate-dependent parameters, several avenues for future research remain. First, while we chose to transform raw actigraphy counts for this analysis, there is growing interest in discrete approaches that explicitly respect the integer-valued nature of the data. Incorporating discrete probability models (e.g., Poisson or Negative Binomial) in combination with structured priors, such as the $\ell_1$-ball projection prior, may further improve interpretability and reduce biologically implausible predictions. This would, however, require methodological innovations in accommodating discrete outcomes within the current Bayesian inference scheme.
	
	Second, we adopted partial pooling strategies for key hyperparameters to account for between-subject variability. Future work could extend these hierarchical assumptions to other aspects of the model, thereby further enhancing the ability to borrow strength across individuals. Along a similar line, more flexible link functions for phase parameters may be explored to ensure parameter constraints are met without compromising sampling efficiency.
	
	Finally, our current setup models each time point independently, aside from the inherent circadian structure. Many actigraphy studies reveal strong temporal dependencies among consecutive observations \citep[see, e.g.,][]{di2023bayesian}, and incorporating explicit serial correlation or autoregressive components could yield more accurate inferences and predictions. This may be achieved by incorporating additional latent processes or by extending our covariate-dependent framework to capture these correlations directly. We believe these developments will provide deeper insights into circadian patterns and strengthen the applicability of our model across a wide range of wearable and longitudinal data contexts.

\section{Data Availability Statement}
Raw actigraphy data can be made available upon reasonable request. 

% Behavioral and Event-Related Potentials data that support the findings in this paper are available on PsyArxiv [Noe and Fischer-Baum, 2020] at \url{https://osf.io/c7k4s/} (DOI
% 10.17605/OSF.IO/C7K4S).

\section{Software}
Supplementary Material includes \texttt{CALC} - a stan software (and R utilities) implementing the methodology outlined in the paper, accompanied by a comprehensive tutorial designed to guide users through replicating the findings detailed in the article.  Stan files and R utilities are also available in Github at XXX (to be made public upon acceptance).

\section{Statements and Declarations}
The authors have no relevant financial or non-financial interests to disclose. The authors have no competing interests to declare that are relevant to the content of this
article. All authors certify that they have no affiliations with or involvement in any organization or entity with any financial interest or non-financial interest in the subject
matter or materials discussed in this manuscript. The authors have no financial or proprietary interests in any material discussed in this article. VK provides scientific consultation for Enliten AI and the Digital In Vivo Alliance, and acknowledges grant funding from the NINDS (K08NS110924, R01NS131399) and the Mike Hogg Foundation. The authors sincerely thank the Editor, Associate Editors, and Referees for their valuable and insightful comments.

\bibliographystyle{apalike}
\bibliography{Biblio.bib}

\end{document}